# PATTERNS OF PHOTOMETRIC AND CHROMOSPHERIC VARIATION AMONG SUN-LIKE STARS: A 20-YEAR PERSPECTIVE

G. W. LOCKWOOD AND B. A. SKIFF

Lowell Observatory

GREGORY W. HENRY AND STEPHEN HENRY

Tennessee State University

R. R. RADICK

Air Force Research Laboratory, National Solar Observatory[1]

AND

S. L. BALIUNAS, R. A. DONAHUE, AND W. SOON

Center for Astrophysics

[1] The National Solar Observatory is operated by AURA, Inc. under cooperative agreement with the National Science Foundation. Partial support for the NSO is provided by the USAF under a Memorandum of Understanding

ABSTRACT

We examine patterns of variation of 32 primarily main sequence Sun-like stars (selected at project onset as stars on or near the main sequence and color index 0.42≤ *(B-V)* ≤1.4), extending our previous 7–12 year time series to 13–20 years by combining Strömgren *b, y* photometry from Lowell Observatory with similar data from Fairborn Observatory. Parallel chromospheric Ca II H & K emission data from the Mount Wilson Observatory span the entire interval. The extended data strengthen the relationship between chromospheric and brightness variability at visible wavelengths derived previously. We show that the full range of photometric variation has probably now been observed for a majority of the program stars. Twenty-seven stars are deemed variable according to an objective statistical criterion. On a year-to-year timescale, young active stars become fainter when their Ca II emission increases while older less active stars such as the Sun become brighter when their Ca II emission increases. The Sun's total irradiance variation, scaled to the *b* and *y* stellar filter photometry, still appears to be somewhat smaller than stars in our limited sample with similar mean chromospheric activity, but we now regard this discrepancy as probably due mainly to our limited stellar sample.

## 1. INTRODUCTION

More than twenty years ago we began exploring how the brightness of stars similar to the Sun varies and how those variations compare with the total irradiance variation of the Sun. Spacecraft measurements since 1980 span three maxima and three minima and yield consecutive smoothed cycle amplitude values of 0.925, 0.896, and 0.895 W/m$^2$ relative to the mean Total Solar Irradiance (TSI) value ~1366 W/m$^2$ (as given by Fröhlich [2006] at http://www.pmodwrc.ch/pmod.php?topic=tsi/composite/SolarConstant ), consistent with the often stated summation that the Sun varies by less than 0.1% over the solar cycle. TSI varies approximately in phase with several manifestations of solar activity such as the sunspot number, whole disk Ca II H & K emission, He 10830 Å emission, 10.7 cm radio flux, etc. (Foukal & Lean 1988; Willson & Hudson 1988). Whether the range of solar variation has been significantly greater in the past remains a matter of some debate; Lean et al. 1995 estimated that at the time of the Maunder Minimum (1645–1715) the TSI may have been 0.24% lower than presently.

Stellar photometric precision at the level of 0.05–0.10% needed to detect year-to-year variations as small as the Sun's is a tough challenge. Nonetheless, our photometry, which approaches but does not quite reach this precision, provides useful comparisons with the solar example on a timescale comparable with the current solar TSI record. This is the main theme of this paper.

The discovery of cyclic magnetic activity variations in stars from early F to early M-spectral type on or near the main sequence by Olin Wilson (1978) at Mount Wilson Observatory was a seminal event in solar-stellar studies. His 10-year study of the variation of chromospheric emission recorded in the cores of the H & K lines of ionized calcium revealed three main types of variability that we now recognize can be roughly classed by stellar age. Older stars in the sample tend either to vary in a smooth, cyclic fashion or have steady levels of H & K emission. Young, active stars vary strongly but irregularly. Paralleling the final three decades of the Mount Wilson stellar program that ended in 2003–4, measurements of solar whole disk Ca K-line emission at Kitt Peak and Sacramento Peak have now tracked the Sun's disk-integrated magnetic activity and so viewed as a star (White et al. 1998; de Toma et al. 2004; W. C. Livingston, private communication).

A 12-year program of *B, V* measurements of sixteen F0–G8 dwarf stars at the Lowell Observatory, begun by H. L. Johnson in 1955 and continued in support of a long-term planetary photometry program by Serkowski (1961) and Jerzykiewicz & Serkowski (1966), showed that those stars do not vary by more than about 1% rms, and for three of them, the standard deviations of their annual mean magnitudes was only 0.4%. Thus, stars close to the Sun's spectral type appeared to be constant within the uncertainties, a result that stood for two decades. A comprehensive survey of stellar variability across the HR diagram derived subsequently from Geneva photometry confirmed that Sun-like dwarf and subgiant stars are, as a class, extraordinarily stable (Grenon 1993).

When we began our photometric study of field dwarfs in 1984 we were mindful of the earlier Lowell *B, V* results, but we also knew that stars much younger than the Sun had detectable brightness variations of a few percent that were negatively correlated with the variations of chromospheric emission (Radick et al. 1983, 1987, 1995; Lockwood et al. 1984). In some cases photometric rotation periods could be derived, illustrating, just as on the Sun, the persistence of activity complexes over weeks to months.

As our new study proceeded, we quickly discovered that the oasis of stability in the HR diagram centered on Sun-like stars contains many examples that are detectably variable at the

sub-1% level. These included a number of long trusted photometric standard stars (Skiff & Lockwood 1986; Lockwood et al. 1997, "Paper I"), and ironically, several comparison stars selected a priori for our survey. By improving photometric precision compared with that of the pioneering Lowell work of Jerzykiewicz and Serkowski, we learned that the population regarded as stable by Jerzykiewicz and Serkowski contained many variables. Of the five stars common to the two programs, we found only one, the G4V star HD 117176, which appears to be constant.

In Paper I we described the Strömgren *b, y* photometric program, the selection of program and comparison stars, the equipment and experimental design, the criteria for detection of variability, and the error budget. We presented the differential light curves for each pairwise combination of stars for the 12-year interval 1984–1995. There were a total of 41 program stars selected from Wilson's original list of 91 stars, organized into 29 trio or quartet groups each containing a pair of comparison stars of similar color. After three years it was apparent that many of the initial comparison stars were slightly variable, so we expanded most of the initial trio groups to quartets, adding an additional comparison star in hopes of fining, ex post facto, at least two stable comparison stars in each group.

Detection of variability rests on the formal significance of the correlation coefficient between pairs of light curves with one star in common. For a quartet, there are six possible combinations; for a trio there are three. Since the program and comparison stars received equal treatment at the telescope, the statistical test we use is indifferent to whether a particular star is a program or comparison star. Ideally, a program star would be found variable and the corresponding comparison stars stable, but sometimes the reverse occurred, a disagreeable revelation when it happened late in the program. That is the main reason why in this paper our initial list of 41 program stars has shrunk to 32 survivors. Of those, 22 have two good – i.e., stable – comparisons stars and 10 must rely on just one.

In Radick et al.1998 (Paper II), using the same 7–12 year data set from Paper I, we described the patterns of photometric variation of 35 stars (including the Sun) compared with their chromospheric behavior measured by the parallel Mount Wilson HK program (Baliunas et al. 1995, 1998). We set six goals in Paper I: (1) to refine further our prior robust metric of solar and stellar variability; (2) to tie the observed variation to astrophysical timescales; (3) to develop procedures for converting differential measurements of variability to the variability of the individual single stars; (4) to relate stellar variability in *b* and *y* passbands to total irradiance measurements of the Sun; (5) to examine aspect (latitudinal) effects, if any; and (6) to study the relationship between photometric variability and chromospheric activity and its variations.

The HK data have the same seasonal cadence as our photometry but often denser coverage (though not necessarily the same date range within seasons owing mainly to differing weather patterns at Mount Wilson and Lowell). The series of photometric and HK annual means reveal temporal patterns that depend on stellar age and mean activity levels (correlated in old stars, anti-correlated in young stars). Chromospheric activity produces a stronger and more easily detected signal than photometric variability, so sometimes we see the former but not the lattter. In the solar example Ca II K emission varies by 20% over a cycle while the smoothed total irradiance varies by less than 0.1%.

In this paper the period of observations available for Papers I and II is lengthened by the addition of partially overlapping photometry from Fairborn Observatory that extends the time series to as long as 20 years, thereby potentially revealing full 22-year magnetic (Hale) cycle timescale patterns for the first time. This also tightens up the variability statistics considerably.

We will show that most of what we learned from 7–12 years of observation remains true over the roughly doubled long interval. We will discuss the lessons learned with regard to the limitations of this particular technique and consider prospects for the future.

## 2. STELLAR SAMPLE, OBSERVATIONS, AND DATA

There were 34 stars in Paper II, all but five (HD 129333, HD 158614, HD 182572, HD 185144, and HD 203387) selected originally from Wilson's sample of 91 stars. In this paper, we drop HD 203387 (because it is a luminosity class III giant; see Fig. 1 in paper II) and HD 176095 (because it lacks a reliable comparison star). Both were observed for only nine seasons. This leaves 32 stars with 13–20 years of observation available for further consideration here.

Table 1 lists the basic properties of the program stars: spectral types, apparent and absolute magnitudes (calculated from Hipparcos parallaxes), mean S-index (an instrumental index of H & K emission derived from measurements of the Mount Wilson program), and the ratio of chromospheric emission and bolometric luminosity, log $R'_{\rm HK}$, calculated from the mean S-index and $B$-$V$ color (see §2.6). The parameter log $R'_{\rm HK}$, originally formulated by Noyes et al. (1984), locates the Sun among lower main sequence stars in an age-activity sequence. Our sample ranges from log $R'_{\rm HK} = -4.4$ (young stars) to $-5.3$ (old stars), bracketing the solar value $-4.94$. We note that log $R'_{\rm HK}$ was developed for lower main sequence stars; thus, applying the formulation to, for example, the subgiant stars now known to exist in the sample introduces a bias when comparing values of log $R'_{\rm HK}$ throughout the sample. Specifically the 0.1nm exit slit may miss some of the flux in the broader emission cores in a subgiant compared to a dwarf. Further, the term subtracted from the emission flux core that removes the non-magnetic flux ($R_{phot}$) from the measured flux would tend to be too large, and the bolometric flux would be too small. How the values of log $R'_{HK}$ should be adjusted for subgiants has not been addressed.

The photometric variability designations in Table 1 follow the Paper I nomenclature: "V" – variability detected with >99% significance, "v" – variability detected with >95% significance, "C" – no variability. A "V+" rating indicates that the amplitude of variation was >0.005 mag (0.5%). For seven stars, the formal variability designation changed (compared with Paper II), either because the confidence interval boundary on the correlation coefficient expanded due to the increase in the number of seasons observed, or because the pattern of stellar variability actually changed in the added years of observation. We indicate these cases by including the Paper II designations in parentheses in Table 1**.**

Figure 1**,** upper panel, shows the program stars plotted on an HR diagram. Several stars appear to be bona fide sub-giants, as based on the Hipparcos parallaxes. Figure 1, lower panel, shows the distribution of the mean chromospheric emission ratio, log $R'_{HK}$ as a function of $B$-$V$ color, coded to indicate photometric variability, and without correction of the calculation for the few subgiants in the sample.

### 2.1. *Lowell Observatory single-channel b, y photoelectric photometry*

Differential photometry carried out by B. A. Skiff using a dedicated 0.5-m telescope began in March 1984 and ended in June 2000 with data recorded on ~1200 nights. The observing scheme comprised measurements of each of the stars of a quartet or a trio through a single filter, Strömgren *b* or *y* (e.g., Lockwood 1983). Four cycles in the order *y, b, b, y* require a total of 40 minutes of telescope time and constitute a nightly observation. This produces six pairwise sets of differential magnitudes for a quartet (three for a trio) per filter, i.e., star 1-star 2, 1-3, …, 3-4.

Typically, Skiff measured 5–10 groups per night and measured each group 8–12 nights per season over the 16-year duration of the program. Especially interesting stars received more intensive scrutiny (e.g., Skiff & Lockwood 1986). Comparison stars included in each group are nearby on the sky, so that the differential atmospheric extinction is negligible.

Since we used the same telescope, photometer, photomultiplier tube, and *b, y* filters throughout the entire observing program, we have been able to maintain the data on the original instrumental magnitude system save only for a very small slowly changing color term, subsequently applied (Lockwood et al. 1997). This secular correction is proportional to $\Delta(b\text{-}y)$ and compensates a slow drift in the color response of the photometer due perhaps to a slow change in the photomultiplier response plus a known 2-nm redward broadening of the *b* filter. The drift is 3x larger in *b* than in *y*. As an example, an F0–G7 star pair having $\Delta(b\text{-}y) = 0.34$ mag (a fairly large color difference) requires an adjustment of 0.00034 mag/yr in *b* and 0.00010 mag/yr in *y*. For further elaboration of the differential photometry error budget and its uncertainties, including the color correction, see Lockwood (2000).

Purists may look askance at two potential sources of systematic error in the Lowell data: (1) use of seasonal mean extinction coefficients rather than nightly measurements, and (2) interference filters operated at outdoor ambient temperature. We satisfied ourselves, however, that neither non-optimum characteristic of our instrumentation and observing protocol results in detectable error or meaningful bias (Lockwood et al. 1997; Lockwood 2000).

### 2.2. *Fairborn Observatory automated photoelectric photometry*

Fairborn Observatory observations, made with a 0.75-m automated reflector located at Washington Camp near Sonoita, AZ, are similar to those at Lowell (Henry 1999) except (1) the Fairborn photometer cycles through the two filters before moving on to the next star in a group, and (2) Fairborn observations are often made more than 1 hour from transit, while Lowell's are not. Fairborn observations cover the interval 1993–2003.

Total integration times per star on the two systems are almost identical, about 1.5 minutes total in each filter, spread over three cycles per night at Fairborn and four at Lowell. The Fairborn data are transformed to the standard Strömgren photometric system (Crawford & Barnes 1970) with the nightly extinction and yearly mean transformation coefficients obtained.

The Fairborn night-to-night precision, about 0.0012 mag rms, is slightly better than that attainable at Lowell owing to the 50% larger Fairborn primary mirror diameter and consequent reduction in Poisson and scintillation noise. Also, the far greater number of Fairborn observations per season (typically ~50 compared with 8–12 at Lowell) results in the Fairborn annual mean magnitudes being more precisely determined than Lowell's. In comparing the Lowell and Fairborn data in detail (next section), we find that the factor of 3–4 improvement in precision expected of Fairborn annual mean magnitudes compared with Lowell is not fully realized, however, possibly because of intrinsic comparison star variability common to both sets of measurements or other error sources that we have not been able to identify.

### 2.3. *Merging Lowell and Fairborn data*

The experimental design and observing scheme of the Lowell and Fairborn observations are compatible in all important respects, insuring a straightforward merger of the two data sets using up to seven years overlap depending on group. We determined a simple magnitude offset for each star pair to transform the Lowell instrumental differential magnitudes to the Fairborn magnitude scale. The offsets range from 0.01–0.02 mag depending on $\Delta(b\text{-}y)$. For our small-range variable stars, color terms are completely negligible. Except where no Fairborn data exist, all the data in this paper are expressed on the Fairborn scale. As a further check on the stability of the fixed offset values, we performed a regression of the yearly offsets with respect to time to look for significant non-zero slopes. None were found.

For HD 124570 and HD 160346, Fairborn observations began the season *after* Lowell observations ended. Here we simply forced the last annual mean Lowell magnitude to equal the first annual mean Fairborn magnitude. This presumes zero variation in the one-year interval between the two time series regardless of the variation before or after the gap. The statistical impact is negligible.

2.4. *Combining differential magnitudes from multiple comparison stars*

Quartet groups are the predominant configuration in our program, offering three pairwise choices to produce the best two comparison stars. If at least two comparison stars are constant, we can make a statistically robust estimate of measurement error that includes the underlying actual comparison star variability and measurement error. Often the choice is not obvious from inspection of the light curves. Choosing the pair with the lowest rms variation of annual mean magnitudes has been our default strategy, and now that we have 15–20 years of data, the uncertainty of the rms comparison star variation is substantially lower than for the 7–12 years of data available for Papers I and II.

An essential choice involves deciding for each star whether to use one comparison star or two. As in Paper II, we followed a three-step procedure. We tested the variance ratios of the possible pairwise combinations, and whenever those variance ratios were statistically equivalent at the >90% level (via an $F$-test), we used both stars, since two comparison stars rather than one reduces the effect of comparison star variability by a factor √2, a gain in precision not to be lightly discarded. It is clearly advantageous to use two comparison stars whenever possible even if one of them appears less stable than the other. This procedure resulted in the selection of two comparison stars in 23 out of 32 cases. For the remaining 9 cases, we were forced to revert to a single comparison star and adopt a nominal value for the estimate of comparison star variability.

Paper II's analysis, which used two comparison stars as often as possible, forced us to exclude several years of trio-only observations for groups that had been promoted to quartets after the third season of observations. Though shorter, the resulting time series seemed a good tradeoff for being able to utilize two comparison stars. In this paper we found a way to salvage the trio-only seasons by establishing a mean offset between the shorter and longer pairwise combinations, and then creating a short (3–4 years typically) prepended, statistically degenerate segment for the shorter of the two pairwise combinations. Splicing, for example, three years of star pair (1–3) observations into the star pair (1–4) time series means that the three first years of the time series will have ~√2 higher internal rms dispersion, a matter of little consequence when considering the entire 15-year or longer time series. We apply this less than ideal approach sparingly (eleven stars), noting that it could influence the final outcome by increasing the rms dispersion of final values by <10% while increasing the length of the time series by as much as 25%.

Our custom in previous papers has been to average the results in $b$ and $y$ by generating a mean value, $(b + y) / 2$. The small amount of astrophysically interesting information available from keeping the magnitudes separate ($\sigma_b > \sigma_y$ by various amounts depending, presumably, on starspot coverage) disappears into the noise for most of the stars.

2.5. *Comparing the errors of the Lowell and Fairborn observations*

The error budgets of the Lowell and Fairborn observations differ slightly, since the two facilities and their modes of operation are not identical. Surprisingly, the on-sky duty cycle of measurements is commensurate, about 60% for the fully automated robotic Fairborn telescopes and the manually-operated Lowell telescope. This means that the two facilities are affected to a roughly similar degree by sky transparency fluctuations during a cycle of measurement. The precision of a typical single differential observation is ~0.0016 for the Fairborn 0.75-m telescope used for the data in this paper (Henry 1999, Figure 11) compared with a slightly larger value, ~0.0020 mag, for the Lowell 0.5-m telescope (Paper I, Figure 18).

According to Young et al. (1991), atmospheric scintillation noise diminishes as a weak power of the telescope aperture, but this advantage in favor of Fairborn may largely be lost for many star groups that transit near the zenith because the Fairborn observations extend over a greater range of hour angle and lead to measurements made at larger airmass. Assessing the distinction, if any, would require a more detailed comparison than is justified for this paper. Photon-counting errors should be roughly a factor of √2 smaller for Fairborn data for comparable integration times, although under typical observing conditions the distinction at least for brighter stars may be academic because observations are not photon-limited at the milli-mag level. Taken together, however, the two quantifiable factors, scintillation and Poisson noise, indicate that the nightly Fairborn observations should be somewhat more precise than Lowell's despite Lowell's slight elevation advantage (2200 m vs. 1700 m), which reduces Lowell's scintillation noise and extinction coefficient.

In addition, Fairborn observations in principle should be internally more consistent because each night's work is reduced to the *uvby* system whereas Lowell observations rely upon seasonally-adjusted mean extinction coefficients and the putative stability of a raw instrumental system (proven over 30 years, except for the small color term mentioned previously). In view of the rather strict prescription for good photometry presented by Young et al. (1991), which includes two of the present authors as co-authors, it is perhaps surprising that the Lowell observations, which break several of the rules for good photometry, match the attained precision level of the Fairborn measurements within about 20% despite the 50% larger aperture and several technical advantages at Fairborn.

We now leave the question of night-to-night ultimate precision with some questions not fully answered, noting simply that observations through the atmosphere seem to have an inherent noise level near 0.001 mag regardless of differences of instrument design and observing protocol.

Lowell observations ended in 2000 after 17 years and Fairborn observations included here mostly fall into the 7–11 year range (1993–2003) with an overlap of several years. In merging the data we have assured ourselves that no significant artifact of either system has been overlooked. Given that a majority of our program stars and many of the comparison stars are slightly variable, examining the data for systematic effects was not completely straightforward. Necessarily, we relied mainly on the behavior of the most constant stars.

### 2.6. *The Mount Wilson Observatory HK data*

The relative fluxes of the Ca II H & K emission cores reported here are the product of the Mount Wilson HK Project initiated by Olin Wilson in 1968 (Wilson 1978), continued and expanded after his retirement (Vaughan et al. 1978; Duncan et al. 1984; Baliunas et al. 1995, 1998). The primary data product is the instrumental S-index, the ratio of the emission flux in 0.1-nm passbands centered on the cores of the H & K lines of ionized calcium by the fluxes in two, 0.2-nm continuum bands bracketing the emission cores. The counts in each channel are corrected for instrumental and sky background. Three separate measurements of each star yield a nightly mean.

Measurements are also normalized by a nightly standardization factor (Baliunas et al. 1995) based on a standard lamp with high flux, augmented by relatively constant standard stars used to check the constancy of the lamp flux. The standard lamp has varied over time so standard star observations are a necessary part of the normalization procedure. In aggregate, the standard stars provide sufficient precision to check the lamp normalization. However, because a few standard stars have not remained constant over the decades, the composition of the standard stars in the aggregate changed from time to time. An iterative procedure yields an average standard deviation relative to S of 0.8% in the flattest stellar records (R. Donahue 2003, private communication).

Over the decades since the HK Project began, the data have been reprocessed several times to account for fresh appraisals of the circumstances of the instrumentation, in particular the calibration information supplied by standard stars and a calibration lamp. Most recently, the data have undergone a comprehensive review by Donahue. There are small quantitative differences between the 1995 compilation and this one, but differences are usually less than 1–2%. The full data set from 1966, including over 2,000 stars, is currently undergoing another reprocessing, and will be published elsewhere.

The S-index is affected by line blanketing in the continuum regions that increases with *B*-*V* color index, thus biasing comparison of Ca II activity for stars of different color. To make S more useful for astrophysical discussions, the index can be transformed to the dimensionless ratio $R'_{\mathrm{HK}}$ (Noyes et al. 1984), and discussed in Section 2. In this paper, we use S when we display the observed time series ("light curves" of H & K emission) but for intercomparing stars, we will use log $R'_{\mathrm{HK}}$.

2.7. *Variability decisions and plots of b, y, and HK time series*

In viewing the light curves presented in this section, it is useful to remember that we make our decisions about intrinsic variability solely on the basis of a formal statistical test. We calculate the confidence level of the correlation coefficient for pairs of differential time series having one star in common (e.g., 1-2 vs. 1-3). At 95% significance we deem the common star "possibly variable" and at 99% we deem it "variable." In this paper we extend that test by first confirming the absence of variability from two or more comparison stars and then test the time series based on the program star minus the mean of the two comparison stars (e.g.. 1- [mean of star 2 and star 3]) vs. the comparison stars themselves (e.g., 2-3). This assures us in most cases that low-level comparison star variability does not contribute significantly to the composite light curve.

How data are averaged with respect to the intrinsic underlying variability signal defines the degrees of freedom (*dof*) used to assess the confidence level of the correlation coefficient. For example, were we to use monthly averages rather than annual means, the *dof* would be ~4 times larger, leading—most likely—to an erroneously high significance level for detected variability. We are most interested in detecting variability on multi-year to decadal timescales. Therefore, the annual mean magnitudes define the appropriate averaging interval. Although this choice arises more from intuition and experience than mathematical exactitude, we believe our method is sound.

Figures 2–4 present a graphical catalog of the $(b+y)/2$ and HK time series for our 32 program stars plus the Sun. We begin by illustrating solar variability in Figure 2 in the same units (stellar magnitudes) as for the stars, noting that we have converted total irradiance variations to $(b + y)/2$ by applying a scale factor of 1.39 (based on blackbody considerations, cf. Paper II) to the TSI variations. The upper panel displays the Ca II K data and the lower panel shows the equivalent brightness variation. Recalling that the recorded solar cycle minimum to maximum range is typically 0.9 W/m$^2$ or 0.66% (based on averages at cycle extrema) we note that when presented as daily values of visible light stellar magnitudes the smooth upper envelope of solar variability (as opposed to a running mean) has a range of 0.0015 mag, roughly twice the TSI range, mainly because of the 40% difference between total irradiance and visible-light flux. The largest transient dip related to a spot transit lies ~0.004 mag below cycle minimum. Thus, to see the Sun —or, rather, its exact analog—as a star we must be able to record long-term stellar variations at a level of ~0.001 mag.

Figure 3 includes the 23 stars that survived the full span of our program with two usable comparison stars. In each panel the embedded solid line is a cubic spline fitted through the annual mean values. The upper panel shows the S-index. The middle panel shows photometric brightness variation of the program star minus the mean of the two comparison stars. The bottom panel shows the light curve of one comparison star minus the other, scaled vertically by a reduction factor $1/\sqrt{2}$ to represent the impact of comparison star variability on the program star light curve.

Figure 4 shows light curves for the 9 stars having only one suitable comparison star. The middle panel is the light curve of the program star minus the chosen comparison star. The bottom panel shows the light curve of one comparison star minus the other plotted with the same vertical scaling as the middle panel and included merely to illustrate graphically what our statistics have already told us, namely that at least one of the stars in the comparison star pair is no good.

Rarely, one of the two selected comparison stars may exhibit statistically significant low-amplitude variability in the matrix of correlation coefficients, though not to the degree required to fail the F test criterion used to choose between adopting one or two comparison stars. Sometimes this occurs by chance, even at the 95% level where there is still one chance in twenty of an accidentally significant correlation. In five such cases we granted the benefit of the doubt, retaining both stars to permit the exact variance arithmetic critically needed to estimate the variability of the program stars and to benefit from the $1/\sqrt{2}$ noise reduction.

An example is HD 18256. The comparison star HD 17659 (star 3) may be slightly variable according to the formal statistic, producing significant correlation between the program star time series (star 1 - [mean of star 2 and star 3]) and the comparison star time series (star 2 - star 3). In such cases we sought a second opinion from the better of the two comparison stars. For HD 18256 we obtain $\sigma_{1\text{-}2,3} = 0.0015$ and $\sigma_{1\text{-}2} = 0.0013$. The “variable” comparison star gives $\sigma_{1\text{-}3} = 0.0019$. Obviously in this example there is a slight penalty in using both comparison stars, but the $1/\sqrt{2}$ noise reduction gained by averaging two comparison stars is some compensation, and exact knowledge of the comparison star variance is desirable (see next section).

We recognize that our work necessarily includes a few *ad hoc* decisions of this type, but in our limited sample we accept less-than-perfect data in order to achieve long records.

## 3. PHOTOMETRIC RESULTS

### 3.1. *Derived intrinsic variability of the program stars*

Our goal is to arrive at a robust estimate of the *intrinsic* photometric variability of each program star (as distinct from the *observed* variation of a program star minus either (a) the mean of the two comparison stars (Figure 3), or (b) minus a sole comparison star (Figure 4). It is important to recognize that intrinsic variability is a derived quantity that involves uncertainty, especially as we approach the limit of detection where comparison star variability is a significant source of noise.

As in Paper II, we perform variance arithmetic on the various pairwise combinations to arrive at a final number for each program star. The principal assumption implicit in this step is that the distributions within the various data sets are approximately Gaussian. As a practical matter, this assumption only matters when the program star variability is less than ~2 times greater than the variability of the comparison star.

The estimated variance of a program star (labeled here "star 1") for which we have two suitable comparison stars (star 2 and star 3), as given in Paper II is:

$$s_1^2 = \sigma^2_{1,23} - ½\ \sigma^2_{2,3} - \varepsilon^2$$

where $s_1^2$ is the calculated estimate of the intrinsic program star variance, $\sigma^2_{1,23}$ is the observed variance of the program star ("1") minus the mean of the two comparison stars (in this case "2" and "3"), $\sigma^2_{2,3}$ is the observed variance of the comparison star pair, and $\varepsilon^2$ is an estimated noise variance.

The first two terms on the right hand side are unbiased measured quantities that nevertheless include some uncertainty since they are based on a data series of *n* points (the number of years of observation) and the assumption of Gaussian behavior. The final term, $\varepsilon^2$, is an estimate of measurement noise based on the lower bound of the distribution of comparison star variances (Paper II, Fig. 4). The value we adopt is $\varepsilon = 0.0006$ mag; it is the same for Lowell and Fairborn observations. We suggest that this represents measurement error when there is no sensible variation in the comparison stars, but we have know way to know exactly why this value falls where it does. The rightmost two terms in this equation become similar when $\sigma_{2,3}$ approaches ~0.001 mag. In our sample this situation rarely occurs, so the exact value adopted for $\varepsilon$ is usually not critical to the determination of $s_1^2$. Second, if the sum of the rightmost two terms is smaller than about ¼ of $\sigma_{1,23}$, precise values of those terms are also relatively unimportant. What this means in practice is that variability is easy to evaluate for active stars, but as we move toward the limits imposed by noise, uncertainty increases. This is a fundamental limitation of differential photometry.

With regard to the approach to pure Gaussian behavior as *n* increases, one might argue that a data series 20 years long (this paper) offers only √2 improvement over one 10 years long (Paper II). Indeed, the results are reassuringly similar for most of our stars. However, we were spurred onward after a decade had passed by the solar example where cycle lengths range from 8 to 14 years. On that basis, we claim that extending the observations for another decade provides substantial assurance that we have now seen the full range of variation for many, if not most, of the stars in our sample. Further, had any subtle instrumental problems (e.g., long term drift) been hidden in the shorter time series, they would have been more fully revealed. In §3.3 we explore the implications of our assumption of pure Gaussian behavior with a fixed background noise level using Monte Carlo simulations.

Table 2 gives the detailed results for 32 program stars and their respective comparison stars. Columns on the left identify the program star and its adopted comparison stars and give the number of years of Lowell and Fairborn observations, the number of years of overlap, the number of years of pre-pended Lowell observation, and the total span of the data. Columns on the right list rms dispersion values for *b, y,* and (*b* + *y*)/2. As noted above, $\sigma_b > \sigma_y$ most of the time. The four rows of data for each stars include, respectively, (row 1) observed dispersion of program star minus mean comparison star; (row 2) observed dispersion of one comparison star minus the other (in parentheses), or, in the case of only one comparison star, a lower limit estimate (in brackets); and (row 3) net intrinsic program star rms variation calculated according to the formula given above. The quantity in brackets is a lower limit estimate, namely ε. The effect of using a lower limit here is to maximize the net variation of the program star. The real variation of those nine stars could be lower than indicated but not higher.

Sometimes the number of years of comparison star data is fewer than the number of years of program star data, as indicated in column 8; this usually has little significant impact on the validity of the variance arithmetic. Finally, for comparison with earlier results based on a shorter time series, row 4 for each star specifies the number of years of observation and corresponding net rms program star dispersion values published in Paper II. Examination of the final columns of rows 3 and 4 for each star shows that in most cases the additional years of observation did not change the final program star rms variation very much (<0.0005 mag for 14 stars and <0.0010 for 19 stars).

Table 2 includes results of two distinct types: (1) the 23 stars (Figure 3) for which we have an exact calculation of intrinsic program star variance based on measured comparison star variance, and (2) the 9 stars (Figure 5) with only one good comparison star, for which the estimated comparison star variance, a lower limit = ε, likely under-estimates the true comparison star variance, thus making the estimated program star variance an upper limit.

As mentioned previously, the decision about whether to use one or two comparison stars is sometimes ambiguous. Table 3 gives particular σ values for five low-amplitude stars where a second opinion based on the more quiescent comparison stars is worth considering. In every case the net program star variance derived from the more quiescent of the two comparison stars is only slightly smaller than the net variance derived from the two comparison stars taken together. So, as a practical matter, at least in these few examples, whether we choose one comparison star or two, the outcome is essentially the same.

3.2. *Have we observed the full range of photometric variability?*

Extending our program has involved a considerable expenditure of telescope time and labor. Has the extra work led to new knowledge? One way of assessing the state of possibly diminishing returns involves graphing the annually accumulating values of the net rms dispersion and peak-to-peak amplitudes for the *n* years of observation, year by year. If trends in dispersion and amplitude flatten out, we may consider that our work is finished, whereas if they increase (or less likely – decrease), then we are adding new information. Our stellar sample divides roughly into thirds – one third with σ increasing (full range possibly not yet observed), one third with essentially constant σ, and one third with σ decreasing (full range observed early in the program). Overall, the effect of additional observations is to put a slowly rising floor under the distribution of net variances plotted, for example, as a function of log $R'_{HK}$. This is a crude diagnostic because it also includes the embedded comparison star variability; nevertheless it does tell us that in a majority of cases, we may as well stop observing, and that twenty years is sufficient for characterizing variability in many of the stars in our sample. A flag in **Table 1** (column 9) indicates which of the three circumstances applies to each star: "+" means σ increasing, "-" decreasing, and "=" no trend.

3.3. *Simulated data*

As a check on the prejudices we have planted in the reader—and ourselves—concerning the reliability of our stated variability and precision values, we experimented with artificial Gaussian data intended to mimic actual observations by having program and comparison star variances in the range we observe. As a test example, we assumed two situations: a program star with intrinsic rms variation of 0.0010 mag (near the lower limit of detection on our program), or a somewhat more easily detected value, 0.0014 mag. We then generated Gaussian comparison star noise variations at five typical levels: $\sigma_{2,3}$ = 0.00035, 0.0005, 0.0007, 0.0010, and 0.0014 mag, the last of these being most common in our program. The corresponding assumed values of $\sigma_{1,23}$ (from the combined intrinsic program star and comparison star noise variances) are therefore 0.00122, 0.00127, 0.00136, 0.00154, and 0.00182 mag, respectively, for the 0.0010 mag intrinsic variability case and 0.00154, 0.00156, .00160, 0.00168, and 0.00182 mag, respectively, for the 0.0014 mag intrinsic variability case. These are the mean observed dispersion values we would obtain if the program star's variation was exactly 0.0010 or 0.0014 mag rms. Then, using a random number generator to produce normally distributed artificial data with zero mean and the above dispersion values we calculated the quantity $s_1^2$ according to the equation in section 3.1 for 800 runs of length 7, 10, 14, and 20 years. In this simulation, we adopt the value $\varepsilon = 0.0006$, noting however that this value is surely not a constant, but yet another normally distributed variable.

Figure 5 gives the range of outcomes, expressed as medians and quartiles of the distributions of calculated net program star variances resulting from the assumed Gaussian distributed inputs. The vertical scale expresses variance, but since what we seek to visualize is the distribution of the standard deviation, we marked the scale with horizontal lines indicating 25% over- and under-estimates of the true variance. The upper quartile, median, and lower quartile points are connected to show trends with increasing time.

We note two significant features of the diagrams. First, the median underestimates the true mean by less than 10% for decade-length time series but approaches the mean even closer as the time series double in length (typical of our data). This is a consequence of the offset in the variance arithmetic due to the $\varepsilon = 0.0006$ mag constant term; it underscores the importance of knowing the value of $\varepsilon$ exactly. The origin of the adopted $\varepsilon$ is described in Paper II, §4.3.1 and Fig. 4 . It seems well defined in our small sample.

Second, there is a systematically greater tendency to under-estimate rather than to over-estimate the true intrinsic variability, owing also to the offset. The good news is that for the 0.0014-mag intrinsic variability case, the estimated variability appears to be robust (i.e., within ±25%) for time series longer than about 14 years. Even for the much tougher problem of detecting 0.0010 mag intrinsic variability, if the comparison star pair is quiescent at the 0.00035–0.0007 level (innermost three quartile lines), the outcome is still good to ±25% at least half the time.

## 4. PATTERNS OF VARIATION

In this section we address, as in Paper II, the general relationships between chromospheric variations, brightness variations, and mean chromospheric activity with particular attention to locating the Sun among our limited sample of Sun-like stars. The comparison now extends over a timescale that, for most if not all of our stars, includes at least one and often more than one full activity cycle (for those stars whose Ca II record shows cycling). We shall not discuss short term (intra-season) variability since our earlier 7–11 year sample adequately addresses their relationship with chromospheric activity and variation.

*4.1. Chromospheric emission variation versus mean activity*

Figure 6 shows the relationship between long-term chromospheric variations expressed as the rms variation of the dimensionless ratio, log $R'_{\mathrm{HK}}$, as a function of mean chromospheric activity. We note that the Sun lies about 30% above the regression line fitted to all stars, a position that is unchanged from Paper II, as is the location of the regression line itself despite the ~2x longer period of stellar observation. Solar data (Figure 3), now extending over three full sunspot cycles, 1976–2004, show that the solar cycle is quite regular with amplitude ~20% in the intensity of K-line emission.

*4.2. Brightness Variation versus Mean Activity*

Analogous to Figure 6, Figure 7 shows the relationship between long-term photometric variation (*b* and *y* averaged) versus mean chromospheric activity. Stars that we deem variable are indicated by filled symbols. The drop lines show the correction from observed total variance to intrinsic variance (corrected for comparison star variance). Stars for which we had only one suitable comparison star and for which we can only estimate an upper limit of variability are separately indicated on Figure 7 using inverted triangles. Two stars (HD 161239 and HD 216385) had negative net variance values (i.e., comparison star variance > program star variance) and are located arbitrarily on the figure at $2\mathrm{x}10^{-5}$. These stars are excluded from discussion because their net variability is indeterminate.

Again, as we found for the chromospheric variation relationship, the regression line (here fitted to the variable stars only denoted by filled symbols) has not shifted despite additional stars now included in the regression that did not pass our test for variability in Paper II. Because of the longer time series (larger *n*) that drives down the correlation coefficient values needed to attain 95% or 99% significance, three stars (HD 13421, HD 124570, HD 182572) formerly considered constant now pass the formal test for variability. They are classified on the MK system as subgiants (confirmed by absolute magnitudes derived from Hipparcos parallaxes) and have formal values of log $R'_{HK} < -5.0$.

*4.3 Correlations between brightness and chromospheric variations*

Among the most robust patterns of stellar behavior that we have found has been the division between young stars, whose photometric behavior is apparently spot-dominated (leading to a negative correlation with chromospheric variations), and old stars, whose photometric behavior is presumably faculae-dominated like the Sun’s (positive correlation with chromospheric variations). The dividing line falls near log $R'_{HK} = -4.7$ and has not shifted with the longer data series. The main difference is that now the correlations are stronger for a number of stars, illustrated in Figure 8**,** upper panel, compared with the same information derived from the shorter time series of Paper II, lower panel. While the shorter sample had fourteen stars whose correlation significance was “low” (p>0.3), the number in that category has now been reduced to eight. Of the four that benefited from additional observations but still retained a low level of correlation significance (HD 13421, HD 182572, HD 103095, and HD 201091), three retained the same sense of correlation while HD 201091 flipped into consistency with the solar example.

Two stars that violated the segregation by correlation sign in the shorter time series, HD 143761 and HD 124570, now do so more definitely, with HD 143761 now attaining "high" ($p<0.05$) significance. Both, alas, have two usable comparison stars, so we cannot blame the discrepancy on an ill-defined comparison star variance. Both lie, however, near the lower limit of detectable variability and HD 143761 retains its former classification as "constant," so these seem likely to be falsely correlated due to statistical noise rather than evidence of a significant departure from the otherwise well-defined relationship. Perils of small-sample-size statistics forbid further speculation on this point.

*4.4 Correlation slope versus mean activity*

Figure 9 shows for our sample of stars the slope of a regression of their annual mean photometric brightness on annual mean chromospheric variability (represented by the S index) as a function of mean chromospheric activity. The dashed line divides faculae-dominated variability (old stars) from spot-dominated variability (young stars). For stars with $\log R'_{HK} > -5.0$ there is a relatively well-defined increase in the amount of photometric variability relative to the chromospheric variability. Six outliers lie well below the rest, including the unusually active star HD 129333. As before, the nine stars with only one usable comparison star are plotted using inverted triangles.

Left of the Sun's location on this diagram there is considerable scatter, which we attribute mainly to the poorly-known level of photometric activity of these stars rather than to an astrophysically-meaningful effect.

This figure, which we consider a key exhibit in the morphology of stellar variability for the Sun and its analogs, raises an interesting question. Is the Sun's location, just slightly above the dividing line, fixed for historical time or could it shift around a bit? Certainly during the three solar cycles of modern observation, there is nothing to suggest that spot activity could overtake facular activity as the principal component of solar variability. The answer, apart from whatever theoretical ruminations might arise, lies in expanding the sample of stars and pushing down the limits of estimated photometric variability as far as possible. The answer, therefore, lies in the indefinite future.

*4.5 Lessons learned*

In this section we discuss how our results might have been improved had we known in 1984 what we know today. We began our survey of Sun-like field stars in 1984 with the new knowledge that young F7–K2 stars in the Hyades vary at the easily detected level of a few percent (Radick et al. 1983, Lockwood et al. 1984). This was a revelation, since Jerzykiewicz and Serkowski (1966) had shown that stars in this spectral range, if they vary at all, do so at levels below 0.5% on a decadal timescale. The Sun itself, shown from spacecraft observations in 1980 to be a variable star on a timescale of days (Willson et al. 1981), had yet to reveal its minuscule cycle timescale 0.1% variation (Fröhlich 2003a, b).

The challenge, as we perceived it in 1984, was therefore to map out variability downward from the easily detected several-percent range of Hyades dwarfs to whatever level our instrumentation would allow. To be reasonably certain of not coming up empty handed we included a number of young, presumably active stars (based on their log $R'_{HK}$ values) in our sample. These rewarded us almost immediately by showing variability.

A preliminary reconnaissance of our capabilities based on observations of planetary targets (e.g., Lockwood 1978, 1981) had shown that a long term precision of 0.002–0.003 mag rms might be achievable for Sun-like stars. In fact, we actually did substantially better than that, with some comparison star pairs included in this paper being demonstrably constant at 0.0005 mag rms (annual means).

Participants in a late 1980s workshop on precision photometry hosted by Russell Genet in Mesa, AZ, several years after the present survey was underway, identified a number of possible technical improvements (Young et al. 1991). These fell into three broad categories: (1) better instrument design, in particular, temperature control of the photomultiplier and filters; (2) optimized observing procedures that incorporate frequent intra-night extinction measurements; and (3) filter passband optimization. We note in passing that some recommendations cannot be incorporated into an ongoing experiment, such as reducing scintillation and Poisson noise by moving to a larger telescope. Others, for example frequent washing of telescope optics, standard practice at Fairborn (where the telescope mirrors lack cover) but not at Lowell, remain of uncertain value. We direct the reader to the Young et al. paper for a comprehensive review of many possible improvements and describe here only those recommendations implemented in the latest generation of robotic telescopes at Fairborn Observatory (Henry 1999).

At Fairborn, a new dual-channel ($b$, $y$) photometer built by Louis Boyd incorporates temperature control over the filters and yields annual mean differential magnitudes sometimes as precise as 0.0001 mag rm. (~3 times better than our best results in this paper). Part of this improvement owes, without doubt, to better Poisson and scintillation statistics from the larger telescope (0.8-m compared with Lowell's 0.5-m). The contribution of temperature stabilization is less easy to quantify.

By far the greatest improvement arises from the new knowledge that the F0–F5 comparison stars now used at Fairborn are more stable than the late F to early K field stars selected for the Lowell survey. In the early F spectral range, γ Dor and δ Sct stars are sometimes encountered (e.g., Henry et al. 2005), but detection usually occurs early enough to permit timely replacement. In the experimental design of the Lowell program, we consciously chose stars in the same color range as the program targets to avoid systematic color-related errors, but Fairborn's temperature-controlled filters and nightly transformation measurements should largely compensate such problems.

Despite this precaution, however, and with the disadvantage of having to search farther afield for suitable comparison stars—an additional worry because of extinction-related errors—the Fairborn program often encounters variable comparison stars that must be discarded and replaced, sometimes inconveniently far along in multi-year time series. This hazard, alas, is unavoidable. The attrition rate among the comparison stars originally chosen for the Lowell program was about 50%, leaving only half of the original sample of program stars surviving with two good comparison stars for this paper. In view of that sobering statistic, and learning from our experience, our firm policy of promptly discarding and replacing bad comparison stars seems a better approach, especially in view of the far larger sample of program stars from which a few dropouts can be tolerated.

This last mentioned problem is astrophysical and cannot be ameliorated by better instrumentation. By going from two comparison stars (trios) to three (quartets), the odds of having at least one pair of stable comparison stars were improved by a factor of three. Would it be even better to organize our program into quintets or even sextets? The disadvantages are obvious: a sextet, say, reduces the program star fraction of measurement to 1/6 the total observing time: it approaches Young's quandary of spending all of one's time calibrating the system and none doing the program! Also, since it lengthens the duration of a nightly measurement cycle accordingly, it runs the risk of larger errors due to intra-cycle sky transparency fluctuations.

## CONCLUSIONS

Starting with 41 program stars two decades ago (Paper I, Paper II), we conclude our study with 22 surviving stars for which a decade-plus time series and a well-behaved pair of comparison stars permit a robust estimate of intrinsic variability. Ten other stars permit an upper limit estimate only. Our sample, therefore, is quite small, although it has produced a power law relationship (Figure 7) with respect to mean chromospheric activity that appears to be observationally robust and that makes astrophysical sense. The lower end of the power law regression is most problematic because as the intrinsic variability level falls below 0.001 mag rms, the contribution of comparison star variability and a perhaps imperfectly understood error budget become significant. We can therefore imagine that the slope of the power law might change if the left-hand tail were better determined.

The location of the Sun on this diagram is of utmost importance. We plotted its position first disregarding the distinction between total irradiance and monochromatic visual brightness, and then with a calculated correction that attempts to reconcile the two scales (see Paper II). Either way, the Sun still lies well below the regression line, leading to the speculation that it may be a low-activity outlier among its stellar cohort (despite being above average in chromospheric variation observed in the same time interval, Figure 6).

In Paper II, we also raised the question of orientation effects, since we observe the Sun in its equatorial plane, whereas our stars are oriented randomly, with the statistical average sub-observer latitude being about 30°. Further speculation on this point seems premature, first because our sample is too small for us to rely upon on an assumed mean sub-observer mean latitude, second because we don't know (observationally at least) what the Sun's variability would look like out of the equatorial plane, and third because the solar example of spot and faculae coverage and evolution might not be appropriate to the larger population of putative solar twins.

We leave these questions to be answered by the far more extensive survey currently being conducted at Fairborn Observatory (Henry 1999). By using more stable comparison stars than we did, the lower limit of detected intrinsic variability can be pushed down by a substantial factor, which should define the lower end of the power law relation better than we have done here. The H&K time series at Mount Wilson Observatory has now been terminated, so the long series of Ca II comparisons must now depend on results of Lowell's Solar-Stellar Spectrograph program (e.g., Hall & Lockwood, 2004; www.lowell.edu/users/jch/sss/index.php) which can only monitor a fraction of the Fairborn sample.

The alternative is high-precision CCD photometry of clusters, where the stability of the entire ensemble provides the photometric reference. Unfortunately, instrumental problems peculiar to CCD photometry (e.g., high-precision flat fields) have largely prevented this goal from being realized.

ACKNOWLEDGMENTS.

This research has been supported at Lowell Observatory by NSF grant ATM 93-13667 and at Tennessee State University by NASA grant NCC5-511 and NSF grant HRD-9706268. SLB is grateful for funding provided by grants from the Richard C. Lounsbery Foundation and JPL (#1270064).

REFERENCES

Allen, C. W. 1973, Astrophysical Quantities, (3rd ed.; London: Athlone Press)

Baliunas, S. L., et al. 1995, ApJ, 438, 269

Baliunas, S. L., Donahue, R. A., & Soon, W. H. , and Henry, G. W. 1998, in ASP Conf. Ser. 154, Cool Stars, Stellar Systems and the Sun, ed. R. A. Donahue & J. A. Bookbinder (San Francisco: ASP), 153

Crawford, D. L., & Barnes, J. V. 1970, AJ, 75, 978

de Toma, G., et al. 2004, ApJ, 609, 1140

Duncan, D. K., et al. 1984, PASP, 96, 707

Foukal, P., & Lean, J. 1988, ApJ 328, 347

Fröhlich, C., 2003a. Solar irradiance variation, in Proc. ISCS Symp. Solar variability as an input to the Earth's Environment, ESA SP-535, (Noordwijk: ESA)

Fröhlich, C., 2003b. Metrologia, 40, S65

Grenon, M. 1993, in ASP Conf. Ser. 40, Inside the Stars, ed. W. Weiss & A. Baglin (San Francisco: ASP), 693

Hall, J. C., & Lockwood, G. W. 2004, ApJ, 614, 942

Henry, G. W. 1999, PASP, 111, 845

Henry, G. W., Fekel, F. C., & Henry, S. M. 2005, AJ, 129, 2815

Jerzykiewicz, M., & Serkowski, K. 1966, Lowell Obs Bull, 6 (No. 137), 295

Lean, J., Beer, J., & Bradley, R. 1995, Geophys. Res. Lett., 22, 3195

Lockwood, G. W. 1978, Icarus, 32, 413–430

Lockwood, G. W. 1981, in Variations of the Solar Constant, ed. S. Sofia, NASA CP-2191, 219

Lockwood, G. W. 1983, in Solar System Photometry Handbook, ed. R. Genet, (Richmond: Willmann-Bell), 2

Lockwood, G. W. 2000, in Third Workshop on Photometry, ed. W. J. Borucki & L. E. Lasher, NASA CP-2000-209614, 9

Lockwood, G. W., Skiff, B. A., & Radick, R. R. 1997, ApJ, 485, 789 [Paper I]

Lockwood, G. W., et al. 1984, PASP, 96, 714

Noyes, R. W., et al. 1984, ApJ, 279, 763

Radick, R. R., et al. 1983, PASP, 95, 621

Radick, R. R., et al. 1987, ApJ, 321, 459

Radick, R. R., et al. 1995, ApJ, 452, 332

Radick, R. R., et al. 1998, ApJS, 118, 239 [Paper II]

Serkowski, K. 1961, Lowell Obs Bull, 5 (No. 11), 157

Skiff, B. A., & Lockwood, G. W. 1986, PASP, 98, 338

Vaughan, A. H., Preston, G. W., & Wilson, O. C. 1978, PASP, 90, 267

White, O. R., et al. 1998, in Synoptic Solar Physics: Proc. of the Eighteenth National Solar Observatory/Sacramento Peak Summer Workshop, ed. K. S. Balasubramanian, J. W. Harvey, & D. M. Rabin (San Francisco: ASP), 293

Willson, R. C., et al. 1981, Science, 211,700

Willson, R. C., & Hudson, H. S. 1988, Nature, 332, 810

Wilson, O. C. 1978, ApJ, 226, 379

Young, A. T., et al. 1991, PASP, 103, 221

Table I. Program stars

| HD (1) | *V* (2) | $M_V$ (3) | *B-V* (4) | Sp. Type (5) | S-index (6) | log $R'_{HK}$ (7) | Long Term Variability (8) | Trend** in σ (9) |
|---|---|---|---|---|---|---|---|---|
| Sun | -26.7 | 4.83 | 0.65 | G2 V | 0.1783 | -4.895 | V | |
| 1835 | 6.4 | 4.84 | 0.66 | G2.5 V | 0.3420 | -4.445 | V+ | = |
| 10476 | 5.2 | 5.87 | 0.84 | K1 V | 0.1896 | -4.938 | (v)* V | - |
| 13421 | 5.6 | 2.50 | 0.56 | G0 IV | 0.1289 | -5.217 | (C) V | - |
| 18256 | 5.6 | 2.83 | 0.43 | F6 V | 0.1804 | -4.758 | V | = |
| 25998 | 5.5 | 3.87 | 0.46 | F7 V | 0.2755 | -4.489 | V+ | + |
| 35296 | 5.0 | 4.17 | 0.53 | F8 V | 0.2982 | -4.438 | V+ | - |
| 39587 | 4.4 | 4.70 | 0.59 | G0- V | 0.3073 | -4.460 | V+ | - |
| 75332 | 6.2 | 3.93 | 0.49 | F7 Vn | 0.2818 | -4.474 | V+ | - |
| 76572 | 6.3 | 2.75 | 0.43 | F6 V | 0.1476 | -4.917 | (C) V | - |
| 81809 | 5.4 | 2.91 | 0.64 | G2 V | 0.1713 | -4.927 | (v) C | = |
| 82885 | 5.4 | 5.16 | 0.77 | G8 IV-V | 0.2679 | -4.674 | V+ | = |
| 103095 | 6.4 | 6.61 | 0.75 | G8 V | 0.1876 | -4.899 | V | = |
| 114710 | 4.2 | 4.42 | 0.57 | F9.5 V | 0.1991 | -4.759 | V+ | - |
| 115383 | 5.2 | 3.92 | 0.59 | G0 Vs | 0.2951 | -4.486 | V+ | - |
| 115404 | 6.5 | 6.24 | 0.94 | K1 V | 0.4965 | -4.529 | V+ | = |
| 120136 | 4.5 | 3.53 | 0.48 | F6 IV | 0.1886 | -4.742 | V+ | - |
| 124570 | 5.5 | 2.92 | 0.54 | F6 IV | 0.1331 | -5.156 | (C) V | + |
| 129333 | 7.5 | 4.95 | 0.61 | G0 V | 0.5475 | -4.148 | V+ | + |
| 131156 | 4.5 | 5.41 | 0.76 | G8 V | 0.4482 | -4.387 | V+ | + |
| 143761 | 5.4 | 4.18 | 0.60 | G0+Va | 0.1492 | -5.046 | C | + |
| 149661 | 5.8 | 5.82 | 0.82 | K2 V | 0.3327 | -4.613 | V+ | = |
| 152391 | 6.6 | 5.51 | 0.76 | G7 V | 0.3840 | -4.460 | V+ | - |
| 157856 | 6.4 | 2.49 | 0.46 | F6 IV-V | 0.1976 | -4.690 | V+ | = |
| 158614 | 5.3 | 4.23 | 0.72 | G9 IV-V | 0.1590 | -5.023 | V+ | - |
| 160346 | 6.5 | 6.38 | 0.96 | K3-V | 0.2904 | -4.811 | C | + |
| 161239 | 5.7 | 2.82 | 0.65 | G2 IIIb | 0.1360 | -5.180 | C | - |
| 182572 | 5.2 | 4.27 | 0.77 | G7 IV | 0.1486 | -5.093 | (C) v | - |
| 185144 | 4.7 | 5.87 | 0.79 | K0 V | 0.2161 | -4.823 | (C) V | = |
| 190007 | 7.5 | 6.87 | 1.14 | K4 V | 0.6396 | -4.711 | V+ | + |
| 201091 | 5.2 | 7.49 | 1.18 | K5 V | 0.6316 | -4.765 | C | = |
| 201092 | 6.0 | 8.33 | 1.37 | K7 V | 0.9447 | -4.910 | V+ | |
| 216385 | 5.2 | 3.02 | 0.48 | F7 IV | 0.1415 | -5.027 | V | - |

* ( ) = designation from Paper II, if different

** see §3.2

Table 2. Basic results

| HD | Comp 1 | Comp 2 | $n_L$ | $n_F$ | $n_{lap}$ | $n_{pre}$ | $n_{tot}$ | $\sigma_b$ | $\sigma_y$ | $\sigma_{by}$ |
|---|---|---|---|---|---|---|---|---|---|---|
| 1835 | 2488 | 1388 | 16 | 16 | 7 | | 19 | 0.0096 | 0.0082 | 0.0088 |
| | | | | | | | 18 | (0.0006) | (0.0008) | (0.0006) |
| | | | | | | | | 0.0093 | 0.0082 | 0.0087 |
| | | | | | | | 11 | | | 0.0097 |
| 10476 | 10697 | 11326 | 16 | 10 | 7 | | 19 | 0.0017 | 0.0015 | 0.0016 |
| | | | | | | | 19 | (0.0022) | (0.0012) | (0.0016) |
| | | | | | | | | 0.0014 | 0.0013 | 0.0014 |
| | | | | | | | 11 | | | 0.0018 |
| 13421 | 13683 | 12414 | 16 | 10 | 7 | 3 | 19 | 0.0011 | 0.0011 | 0.0011 |
| | | | | | | | 16 | (0.0012) | (0.0009) | (0.0009) |
| | | | | | | | | 0.0006 | 0.0007 | 0.0007 |
| | | | | | | | 9 | | | 0.0004 |
| 18256 | 18404 | 17659 | 15 | 10 | 7 | | 19 | 0.0016 | 0.0016 | 0.0015 |
| | | | | | | | 19 | (.0014) | (0.0010) | (0.0011) |
| | | | | | | | | 0.0014 | 0.0014 | 0.0014 |
| | | | | | | | 11 | | | 0.0013 |
| 25998 | 24747 | 23885 | 13 | | | | 13 | 0.0029 | 0.0027 | 0.0027 |
| | | | | | | | 13 | (0.0018) | (0.0012) | (0.0014) |
| | | | | | | | | 0.0026 | 0.0026 | 0.0026 |
| | | | | | | | 11 | | | 0.0020 |
| 35296 | 33276 | 38558 | 17 | | | | 17 | 0.0049 | 0.0041 | 0.0045 |
| | | | | | | | 17 | (0.0016) | (0.0012) | (0.0013) |
| | | | | | | | | 0.0049 | 0.0041 | 0.0045 |
| | | | | | | | 12 | | | 0.0033 |
| 39587 | 33276 | 38558 | 17 | | | | 17 | 0.0067 | 0.0058 | 0.0062 |
| | | | | | | | 17 | (0.0016) | (0.0012) | (0.0013) |
| | | | | | | | | 0.0066 | 0.0058 | 0.0062 |
| | | | | | | | 12 | | | 0.0063 |
| 75332 | 73596 | 78234 | 10 | 12 | 3 | | 18 | 0.0063 | 0.0054 | 0.0057 |
| | | | | | | | 20 | (0.0012) | (0.0009) | (0.0010) |
| | | | | | | | | 0.0060 | 0.0052 | 0.0057 |
| | | | | | | | 10 | | | 0.0067 |
| 76572 | 73596 | 78234 | 12 | 4 | 3 | | 13 | 0.0008 | 0.0009 | 0.0008 |
| | | | | | | | 20 | (0.0011) | (0.0006) | (0.0008) |
| | | | | | | | | 0.0004 | 0.0008 | 0.0006 |
| | | | | | | | 12 | | | 0.0005 |
| 81809 | 81342 | | 17 | 11 | 8 | 4 | 20 | 0.0014 | 0.0010 | 0.0012 |
| | | | | | | | | | | [0.0006] |
| | | | | | | | | | | <0.0011 |
| | | | | | | | 9 | | | 0.0009 |
| 82885 | 83951 | 83525 | 17 | 11 | 8 | 1 | 20 | 0.0054 | 0.0046 | 0.0050 |
| | | | | | | | 19 | (0.0008) | (0.0006) | (0.0006) |
| | | | | | | | | 0.0054 | 0.0046 | 0.0050 |
| | | | | | | | 11 | | | 0.0033 |
| 103095 | 102713 | | 17 | 11 | 8 | 4 | 20 | 0.0015 | 0.0012 | 0.0013 |
| | | | | | | | | | | [0.0006] |
| | | | | | | | | | | <.0012 |
| | | | | | | | 8 | | | 0.0009 |
| 114710 | 111812 | | 17 | 4 | 4 | | 17 | 0.0026 | 0.0019 | 0.0023 |
| | | | | | | | | | | [0.0006] |
| | | | | | | | | | | <0.0023 |
| | | | | | | | 8 | | | 0.0020 |
| 115383 | 117304 | 117176 | 17 | 11 | 8 | | 20 | 0.0052 | 0.0045 | 0.0049 |
| | | | | | | | 17 | (0.0021) | (0.0020) | (0.0019) |
| | | | | | | | | 0.0051 | 0.0044 | 0.00048 |
| | | | | | | | 12 | | | 0.0055 |
| 115404 | 113848 | | 11 | 11 | 2 | | 20 | 0.0074 | 0.0067 | 0.0071 |
| | | | | | | | | | | [0.0006] |
| | | | | | | | | | | <0.0071 |
| | | | | | | | 7 | | | 0.0026 |
| 120136 | 121560 | 120601 | 17 | 11 | 8 | 6 | 20 | 0.0035 | 0.0026 | 0.0030 |
| | | | | | | | 16 | (0.0006) | (0.0006) | (0.0005) |
| | | | | | | | | 0.0033 | 0.0026 | 0.0030 |
| | | | | | | | 8 | | | 0.0025 |
| 124570 | 125451 | 123845 | 9 | 11 | | 4 | 20 | 0.0013 | 0.0019 | 0.0015 |
| | | | | | | | 16 | (0.0012) | (0.0008) | (0.0009) |
| | | | | | | | | 0.0012 | 0.0018 | 0.0014 |
| | | | | | | | 16 | | | 0.0014 |
| 129333 | 129390 | 131330 | 14 | 11 | 5 | | 20 | 0.0429 | 0.0377 | 0.0402 |
| | | | | | | | 16 | (0.0007) | (0.0009) | (0.0007) |
| | | | | | | | | 0.0429 | 0.0377 | 0.0403 |
| | | | | | | | 11 | | | 0.0238 |
| 131156 | 129972 | 132146 | 14 | | | 1 | 14 | 0.0100 | 0.0088 | 0.0094 |
| | | | | | | | 13 | (0.0025) | (0.0017) | (0.0020) |
| | | | | | | | | 0.0099 | 0.0087 | 0.0093 |
| | | | | | | | 11 | | | 0.0090 |
| 143761 | 142091 | 140716 | 17 | 9 | 7 | | 18 | 0.0019 | 0.0012 | 0.0015 |
| | | | | | 17 | | 12 | (0.0016) | (0.0017) | (0.0013) |
| | | | | | | | | 0.0016 | 0.0009 | 0.0009 |
| | | | | | | | 12 | | | 0.0009 |
| 149661 | 150050 | 152569 | 14 | 11 | 5 | | 20 | 0.0109 | 0.0064 | 0.0084 |
| | | | | | | | 14 | (0.0050) | (0.0024) | (0.0036) |
| | | | | | | | | 0.0073 | 0.0059 | 0.0066 |
| | | | | | | | 12 | | | 0.0067 |
| 152391 | 150050 | 152569 | 14 | 11 | 5 | | 20 | 0.0155 | 0.0134 | 0.0144 |
| | | | | | | | 14 | (0.0050) | (0.0024) | (0.0036) |
| | | | | | | | | 0.0148 | 0.0128 | 0.0138 |
| | | | | | | | 12 | | | 0.0144 |
| 157856 | 156635 | 157347 | 16 | | | | 16 | 0.0017 | 0.0015 | 0.0014 |
| | | | | | | | | (0.0014) | (0.0011) | (0.0007) |
| | | | | | | | | 0.0015 | 0.0014 | 0.0014 |
| | | | | | | | 12 | | | 0.0014 |
| 158214 | 156635 | 157347 | 16 | | | | 16 | 0.0019 | 0.0018 | 0.0018 |
| | | | | | | | | (0.0014) | (0.0011) | (0.0007) |
| | | | | | | | | 0.0018 | 0.0017 | 0.0017 |
| | | | | | | | 12 | | | 0.0019 |
| 160346 | 160385 | 1608/23 | 13 | 4 | | | 13 | 0.0016 | 0.0036 | 0.0020 |
| | | | | | | | 9 | (0.0028) | (0.0022) | (0.0024) |
| | | | | | | | | --- | 0.0008 | --- |
| | | | | | | | 9 | | | 0.0011 |
| 161239 | 160935 | | 16 | 11 | 7 | 1 | 20 | 0.0015 | 0.0013 | 0.0014 |
| | | | | | | | | | | [0.0006] |
| | | | | | | | | | | <0.0013 |
| | | | | | | | 11 | | | 0.0010 |
| 182572 | 180868 | 182899 | 16 | 11 | 7 | 4 | 20 | 0.0014 | 0.0014 | 0.0013 |
| | | | | | | | 18 | (0.0017) | (0.0012) | (0.0014) |
| | | | | | | | | 0.0010 | 0.0012 | 0.0010 |
| | | | | | | | 14 | | | 0.0009 |
| 185144 | 187340 | | 11 | 11 | 2 | 4 | 20 | 0.0020 | 0.0017 | 0.0019 |
| | | | | | | | | | | [0.0006] |
| | | | | | | | | | | <0.0019 |
| | | | | | | | 7 | | | 0.0003 |
| 190007 | 190521 | 190498 | 16 | 11 | 7 | 4 | 20 | 0.0058 | 0.0047 | 0.0052 |
| | | | | | | | 16 | (0.0029) | (0.0017) | (0.0022) |
| | | | | | | | | 0.0056 | 0.0046 | 0.0050 |
| | | | | | | | 7 | | | 0.0047 |
| 201091 | 201154* | | 16 | 8 | 4 | | 20 | 0.0030 | 0.0017 | 0.0024 |
| | | | | | | | | | | [0.0006] |
| | | | | | | | | | | <0.0024 |
| | | | | | | | 11 | | | 0.0013 |
| 201092 | 200031 | | 16 | | | | 16 | 0.0040 | 0.0017 | 0.0030 |
| | | | | | | | | | | [0.0006] |
| | | | | | | | | | | <0.0030 |
| | | | | | | | 11 | | | 0.0037 |
| 216385 | 217232 | | 16 | | | | 16 | 0.0011 | 0.0008 | 0.0009 |
| | | | | | | | | | | ]0.0006] |
| | | | | | | | | | | <0.0005 |
| | | | | | | | 11 | | | 0.0008 |

*201091 group original comp 201154 transformed to new comp 200031

Table 3. Stars with one slightly variable comparison star

| HD | σ (Prog. Star - both comps) | σ (Prog. Star -best comp) | σ(Prog. Star -“variable” comp) | σ (comp.) |
|---|---|---|---|---|
| 18256 | 0.0015 | 0.0013 | 0.0019 | 0.0011 |
| 25998 | 0.0027 | 0.0025 | 0.0031 | 0.0014 |
| 76572 | 0.0008 | 0.0008 | 0.0009 | 0.0008 |
| 124570 | 0.0015 | 0.0014 | 0.0017 | 0.0009 |
| 160346 | 0.0020 | 0.0016 | 0.0019 | 0.0024 |

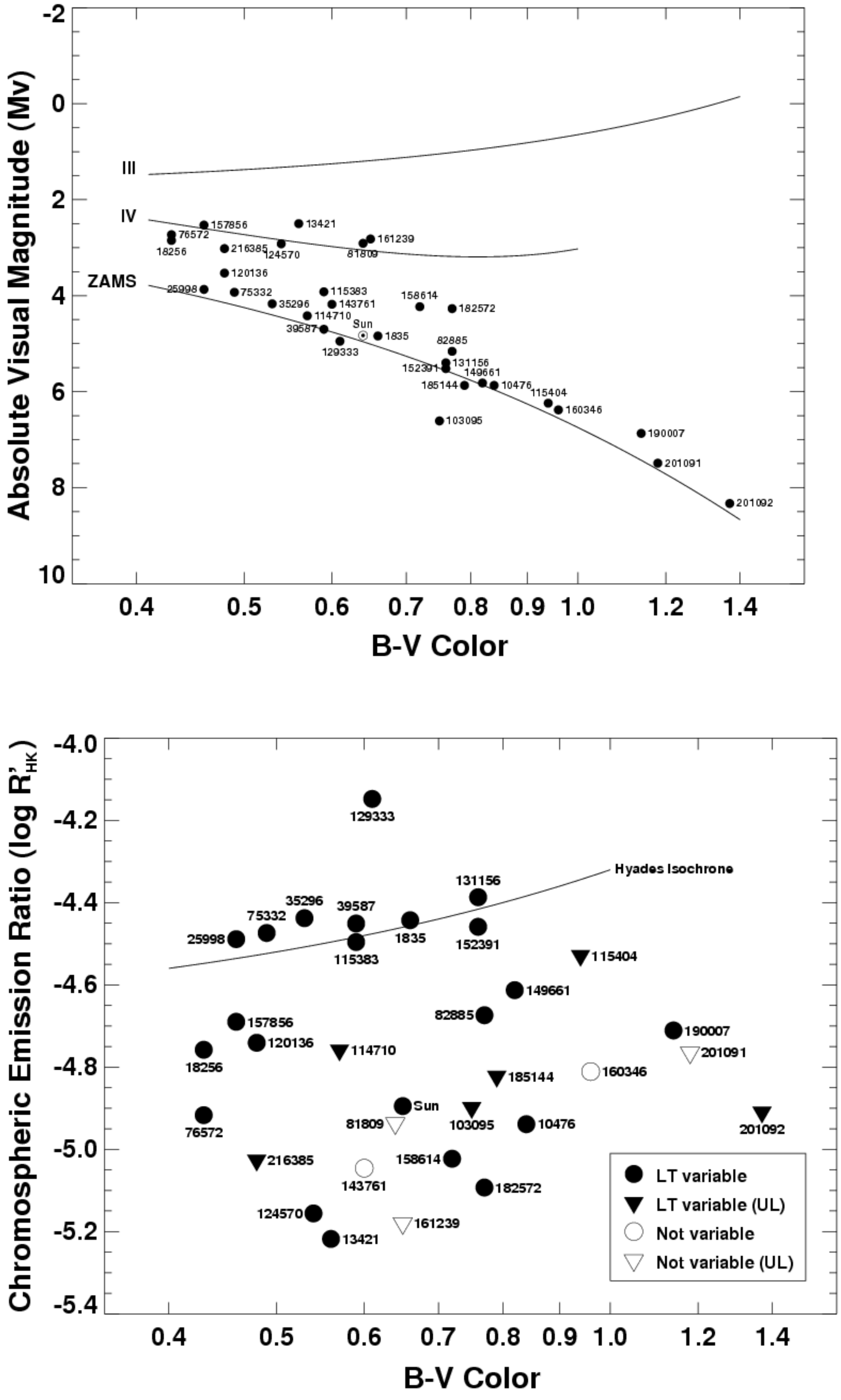

FIG. 1— *upper*: H-R diagram for the stars of our sample. The lines indicating luminosity class III, IV, and ZAMS are based on data from Allen (1973). *lower*: Activity-color diagram for the stars of our sample. Those found to vary photometrically on the long-term timescale are represented by filled symbols.

.

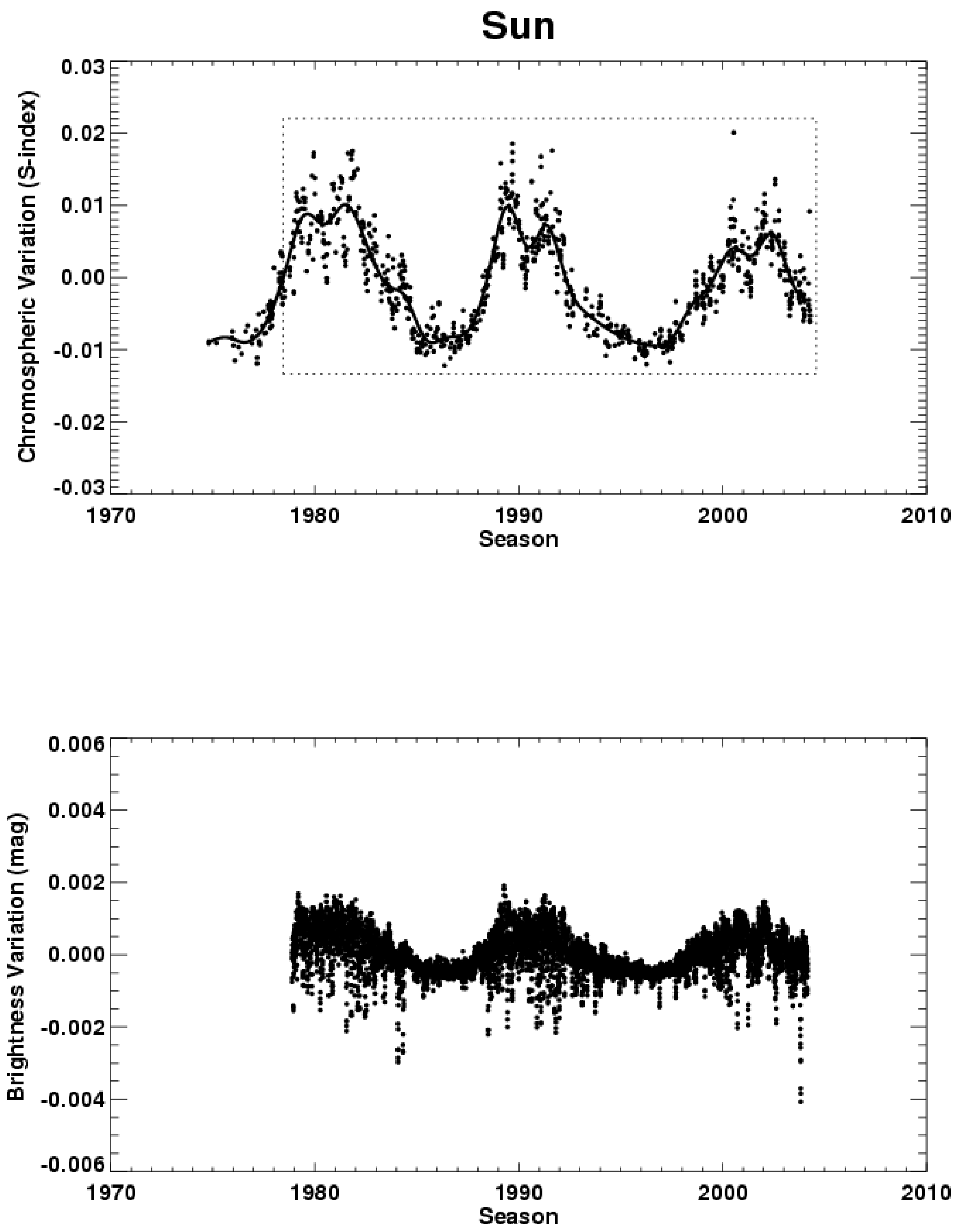

FIG. 2—Ca II K and total irradiance variability of the Sun. The K data are from White et al. 1998 plus updates from Livingston (private communication). The irradiance data (Fröhlich, 2003a,b_ can be found at http://www.pmodwrc.ch/pmod.php?topic=tsi/composite/SolarConstant (graph) or http://www.ngdc.noaa.gov/stp/SOLAR/IRRADIANCE/irrad.html (links to tabular data)

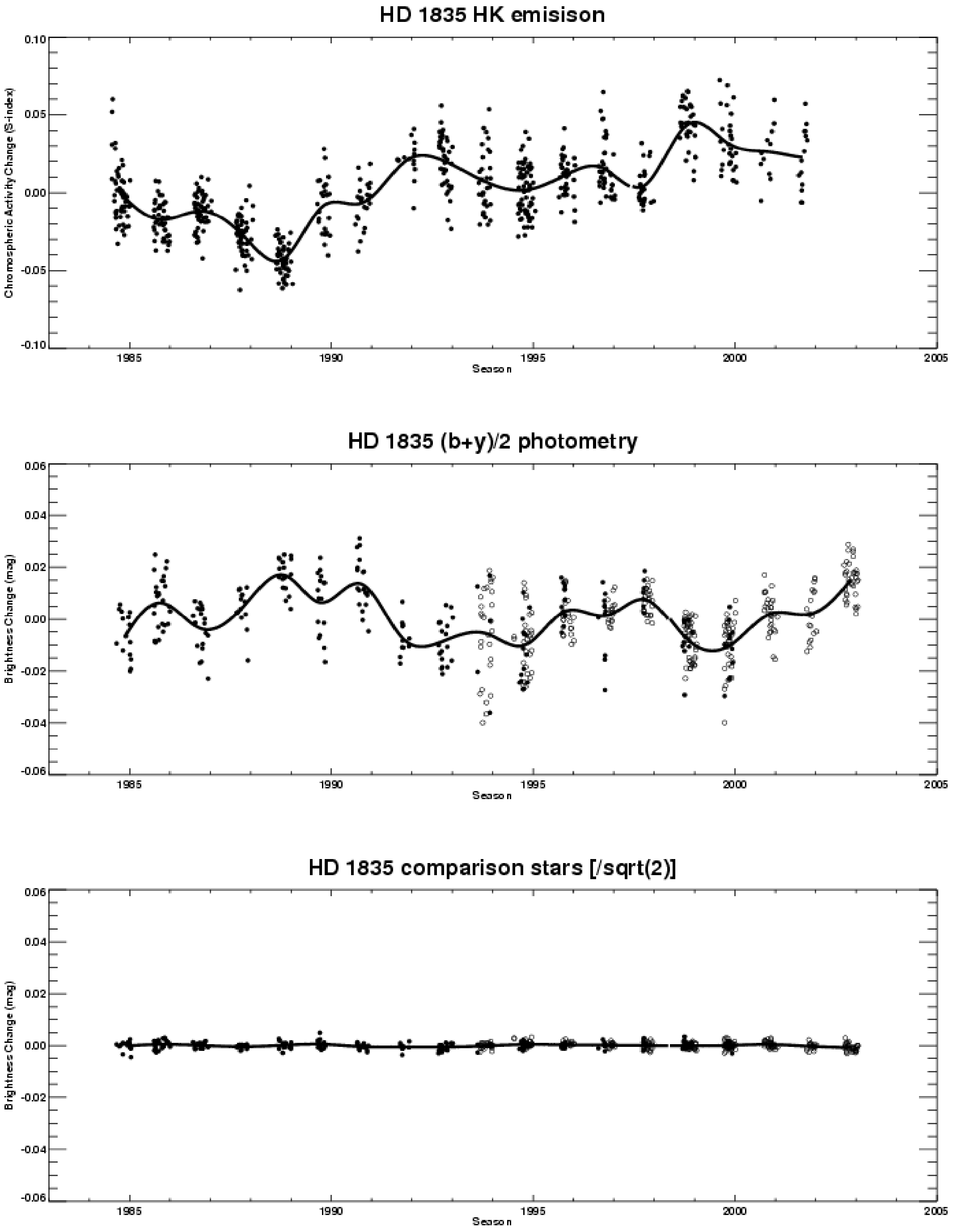

FIG. 3a—HD 1835. Chromospheric Ca II HK emission (*upper*), photometric program star (*middle*) and photometric comparison star (*lower*) time series plots for the stars of our sample having two valid comparison stars. Brightness increases upward in all cases, and the bottom panel is scaled by √2 so that the statistical impact of variability is commensurate in the lower two panels.

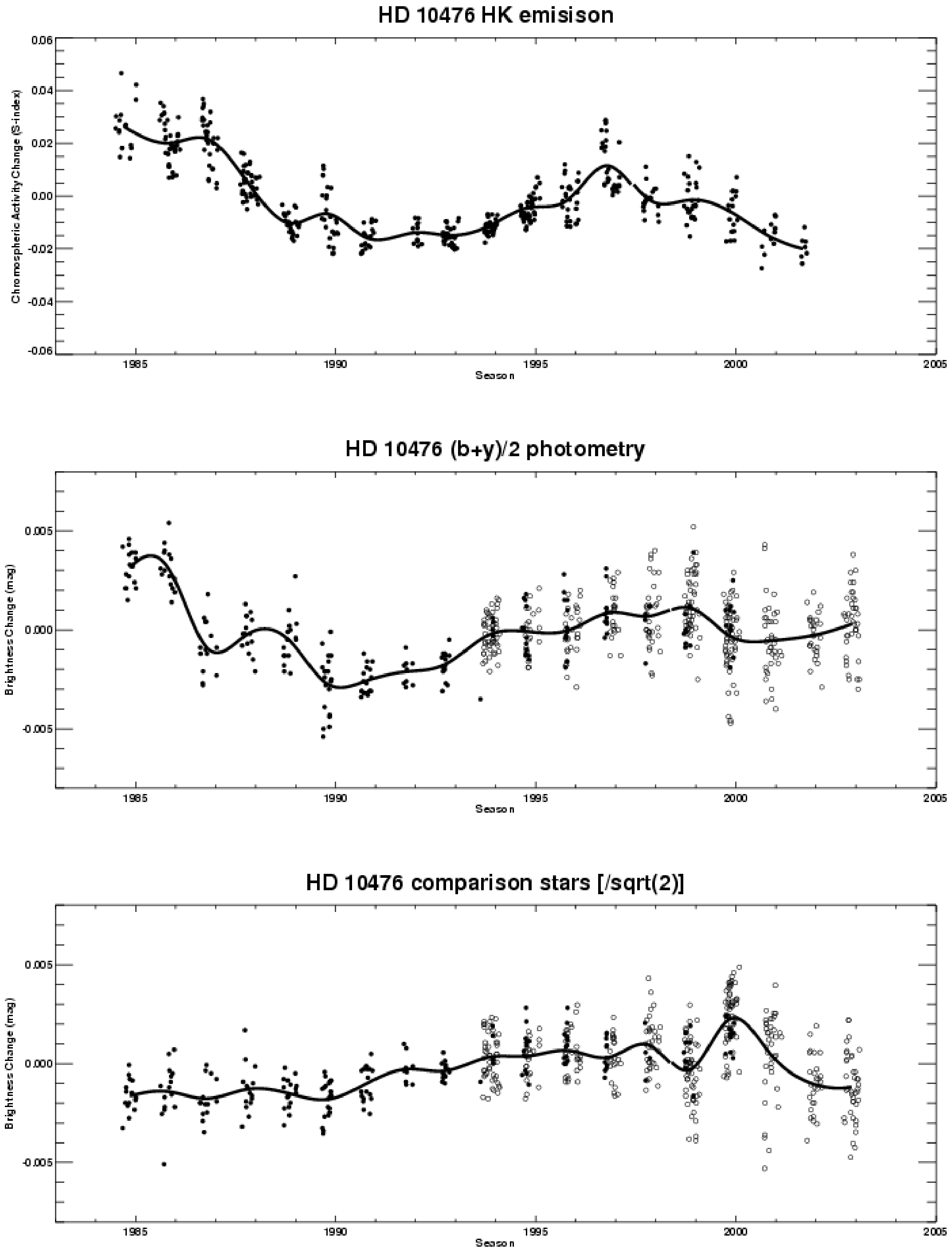

FIG. 3b—HD 10476

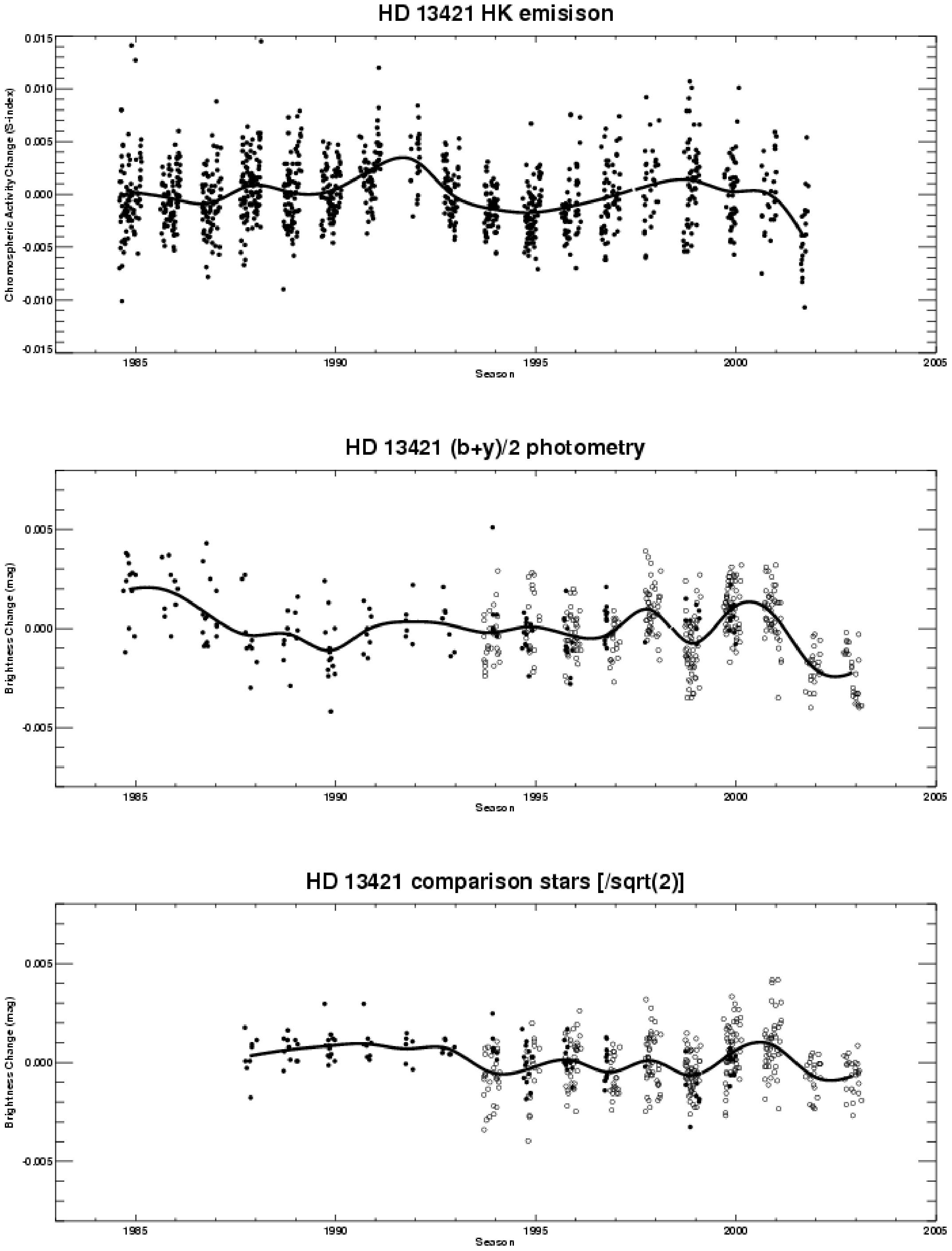

FIG. 3c—HD 13421

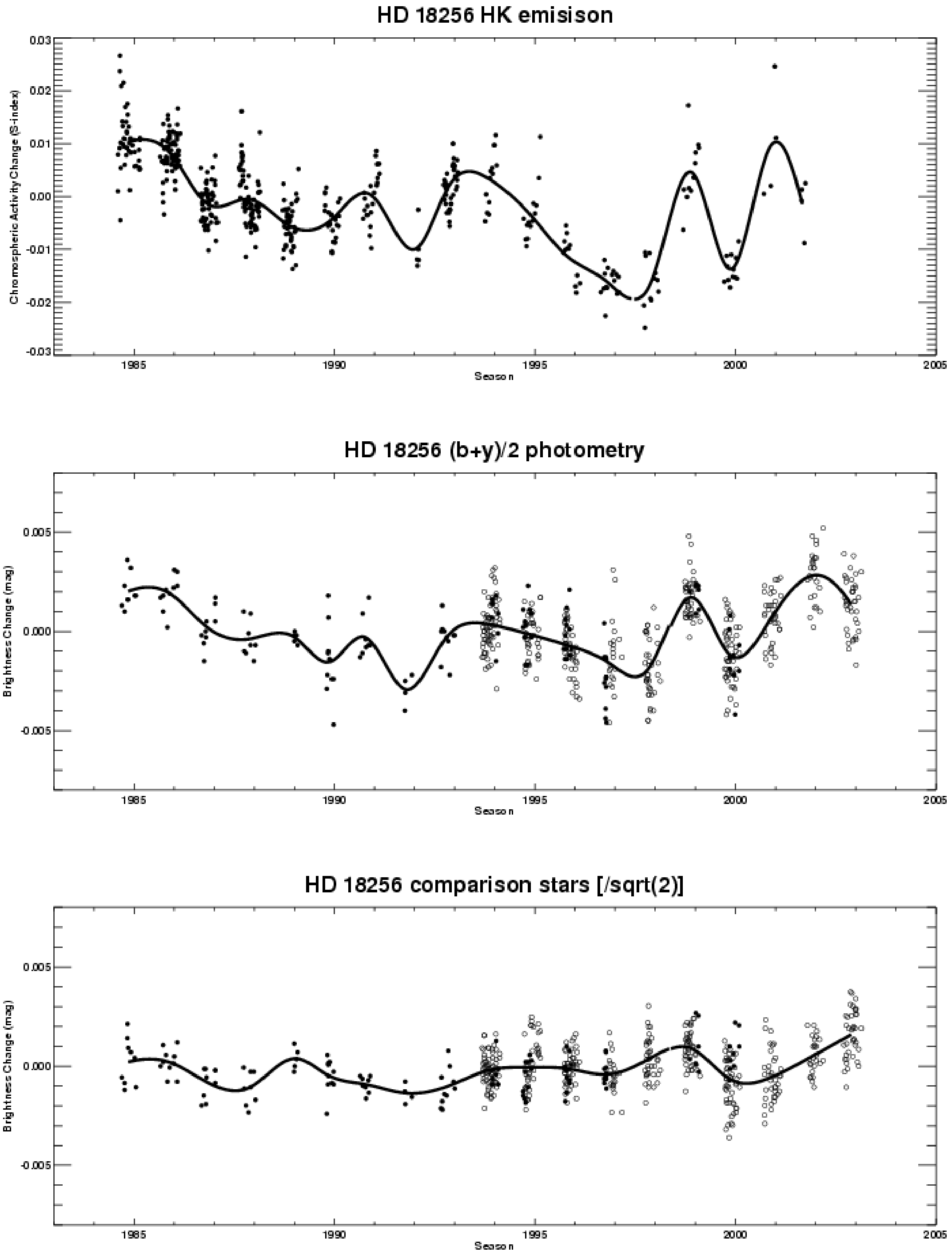

FIG. 3d—HD 18256

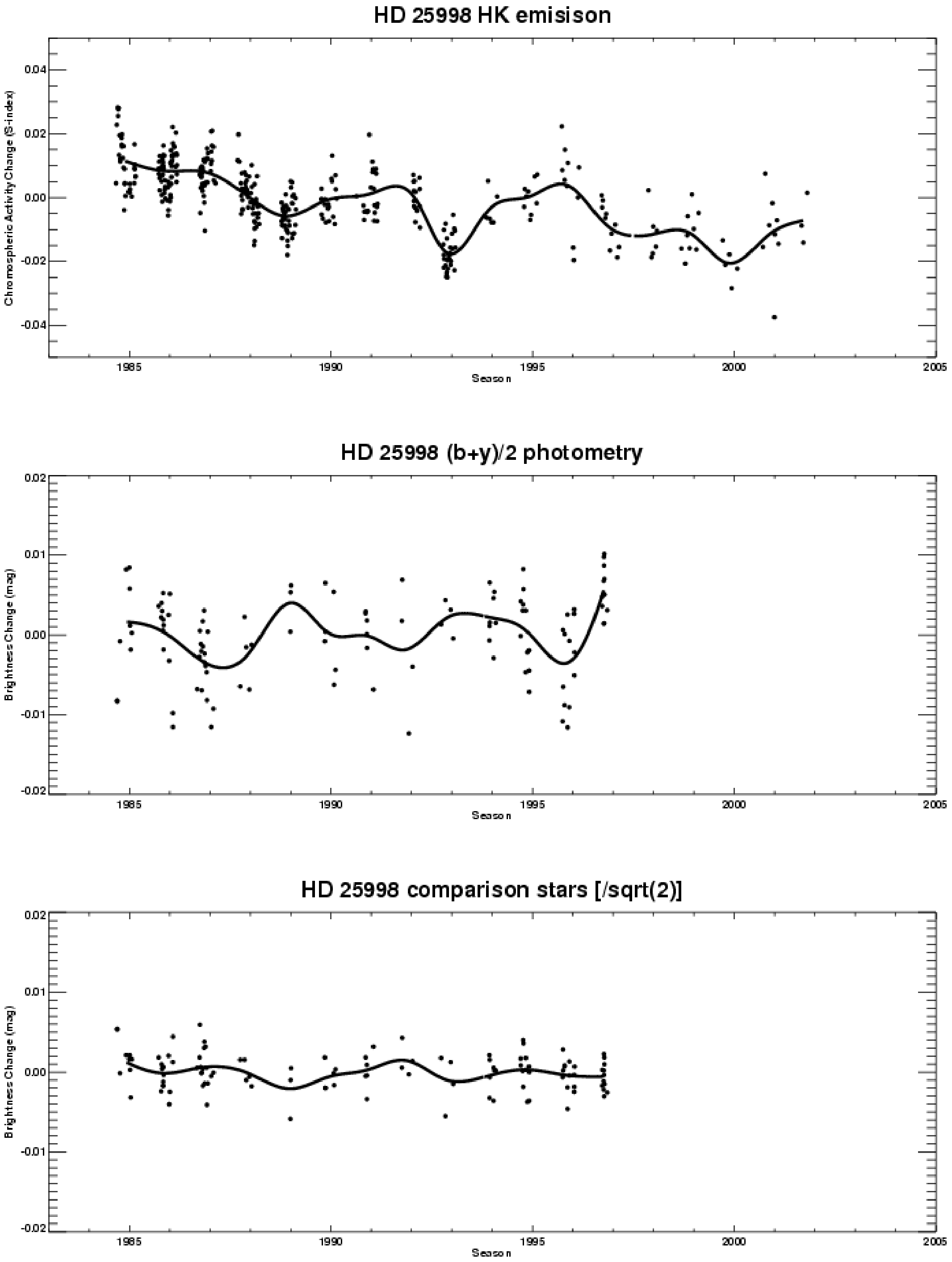

FIG. 3e—HD 25998

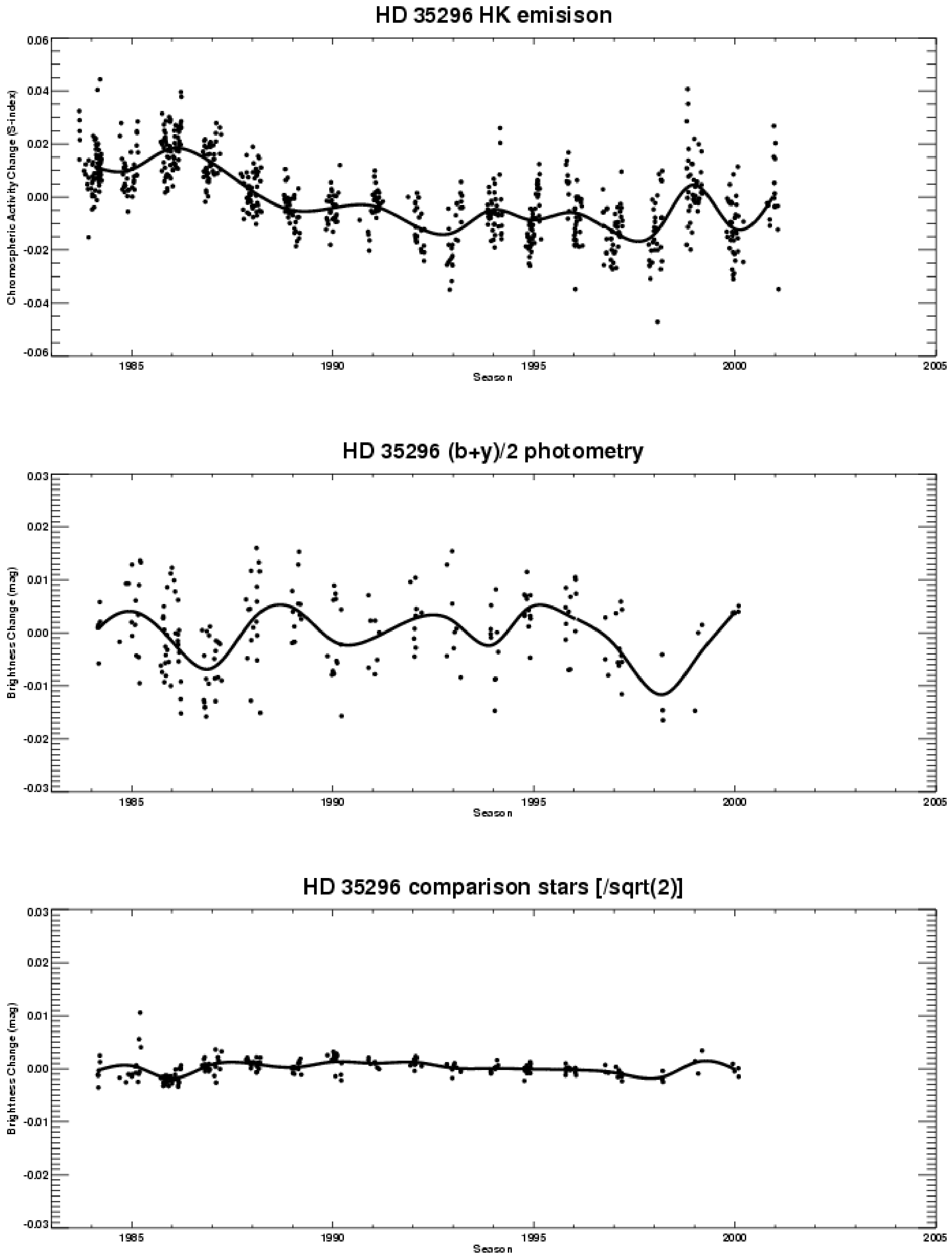

FIG. 3f—HD 35296

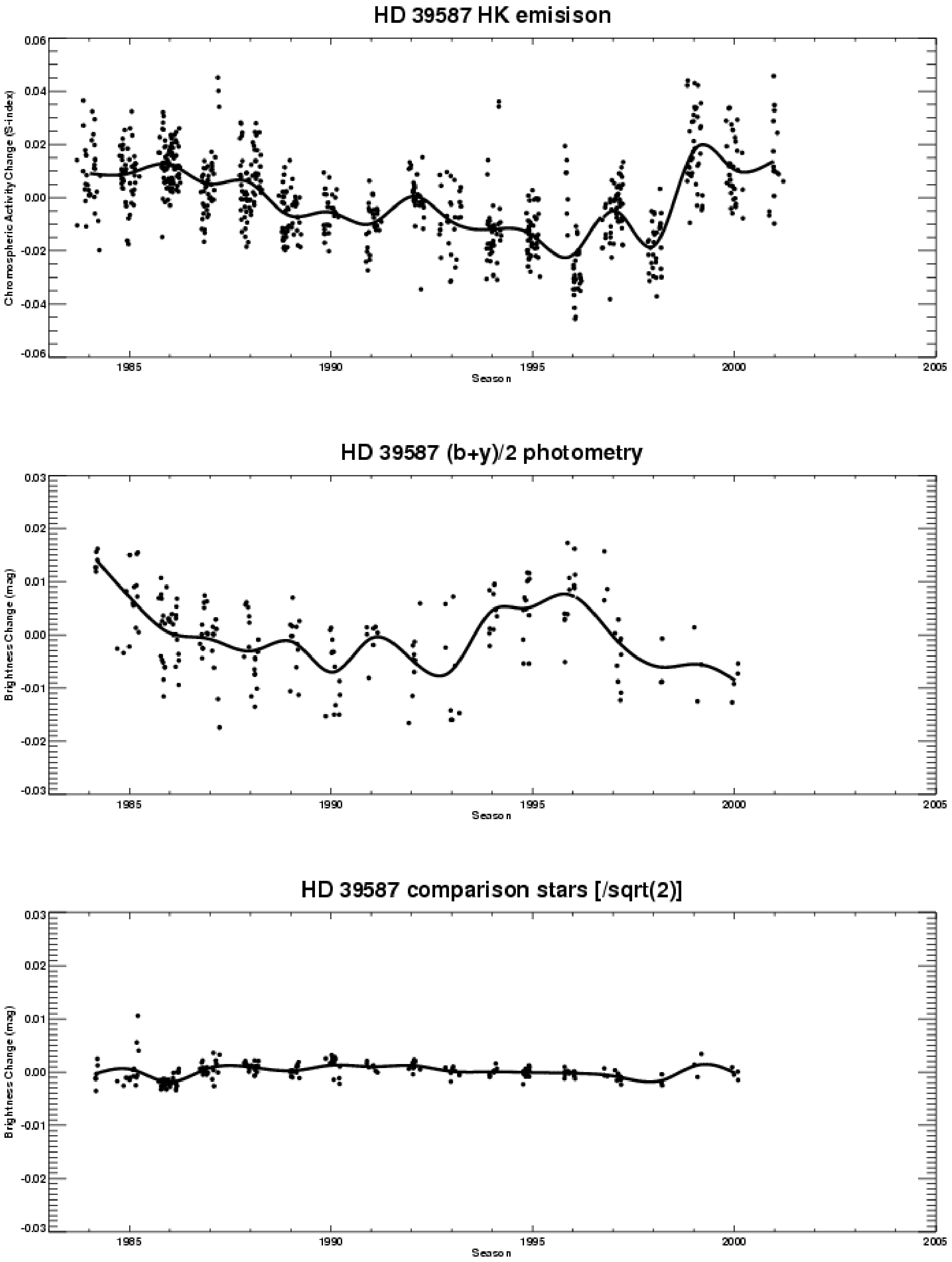

FIG. 3g—HD 39587

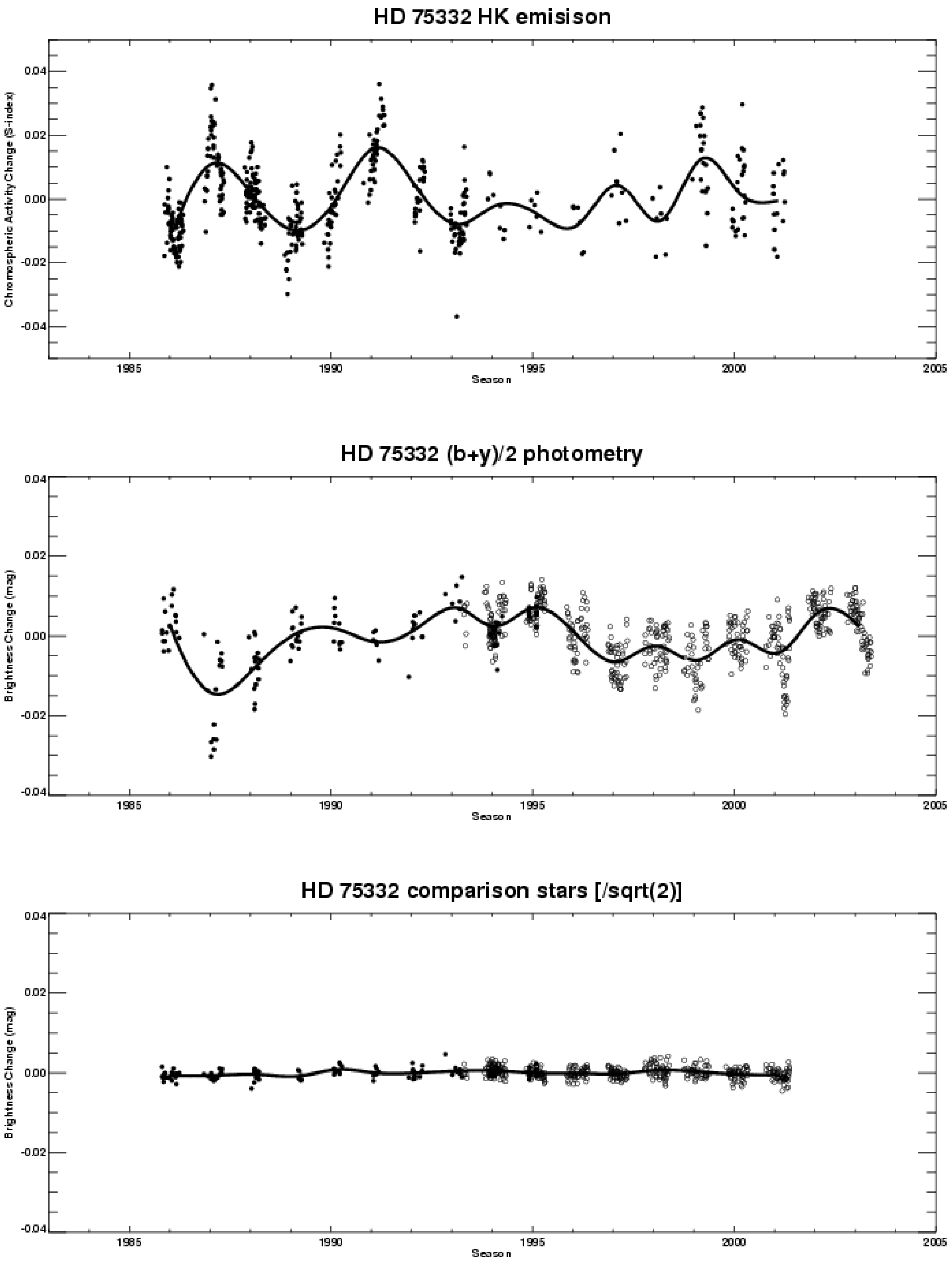

FIG. 3h—HD 75332

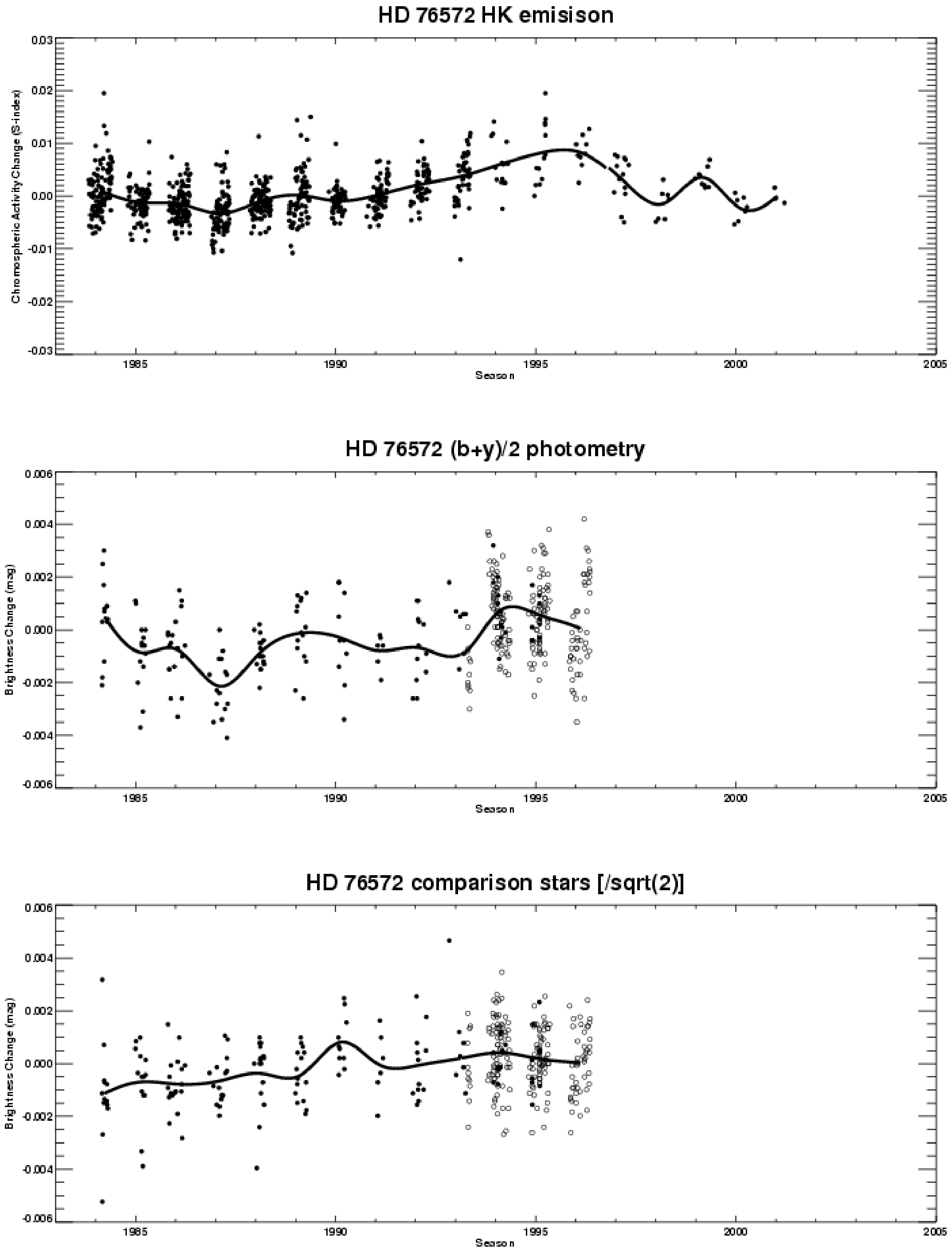

FIG. 3i—HD 76572

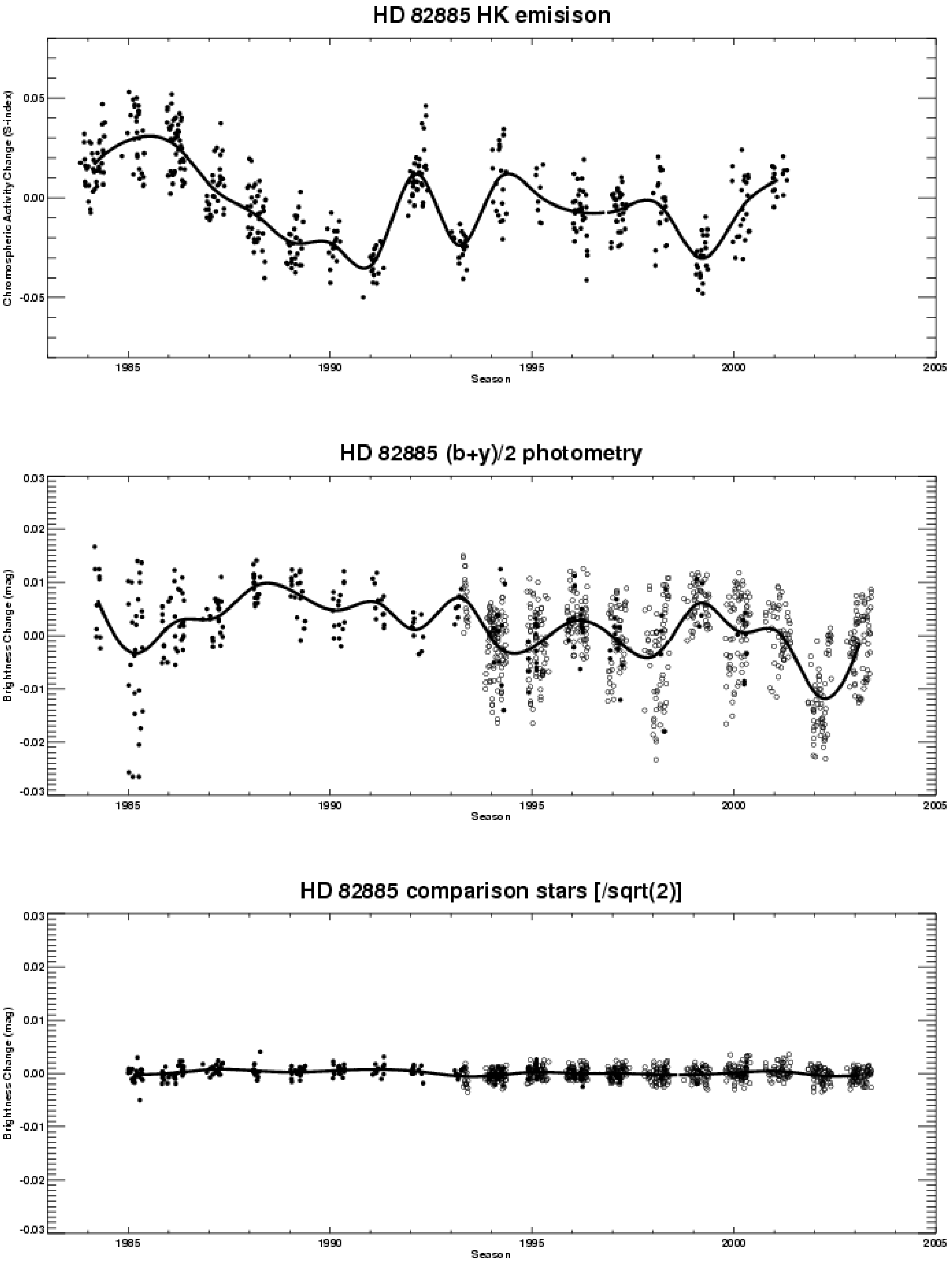

FIG. 3j—HD 82885

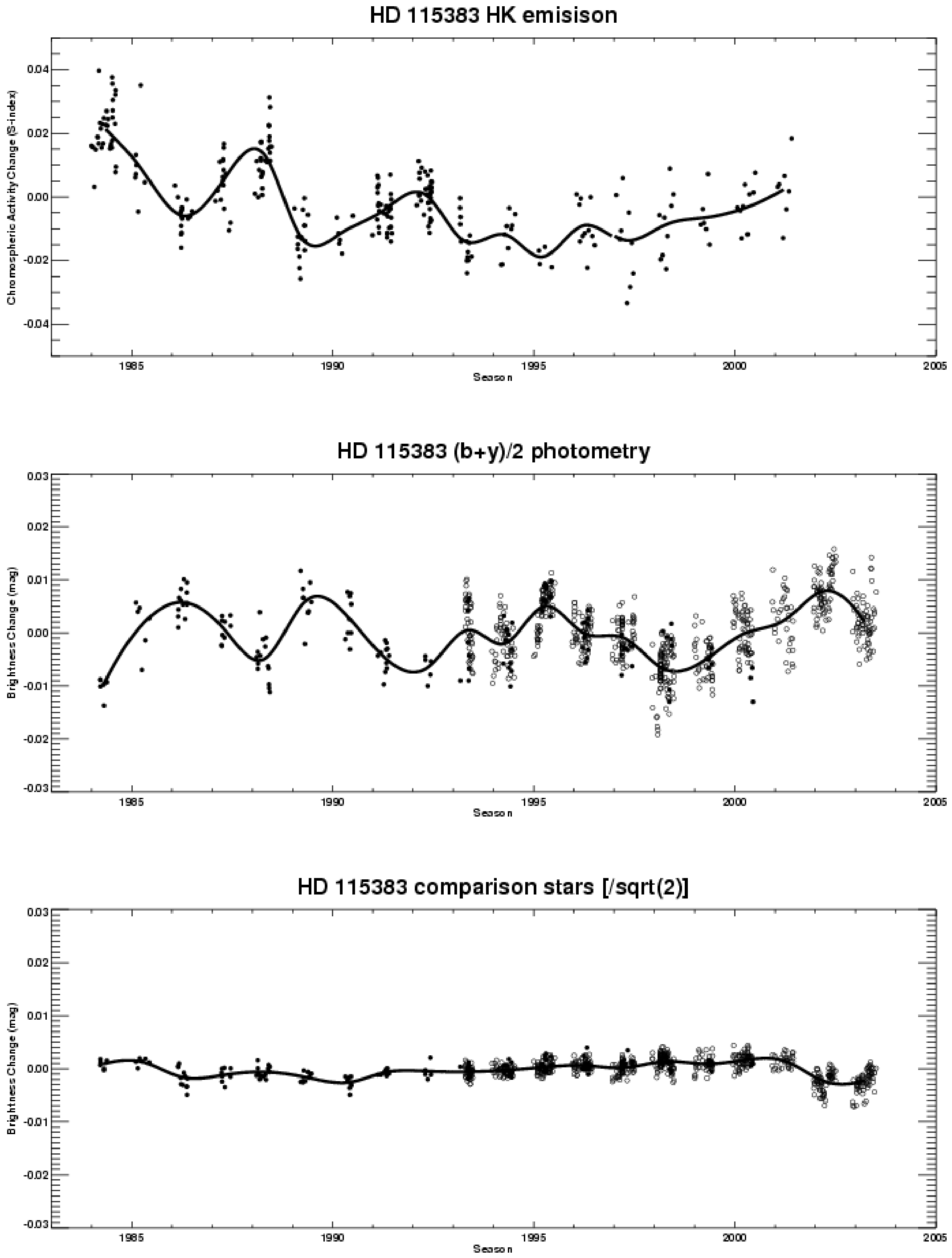

FIG. 3k—HD 115383

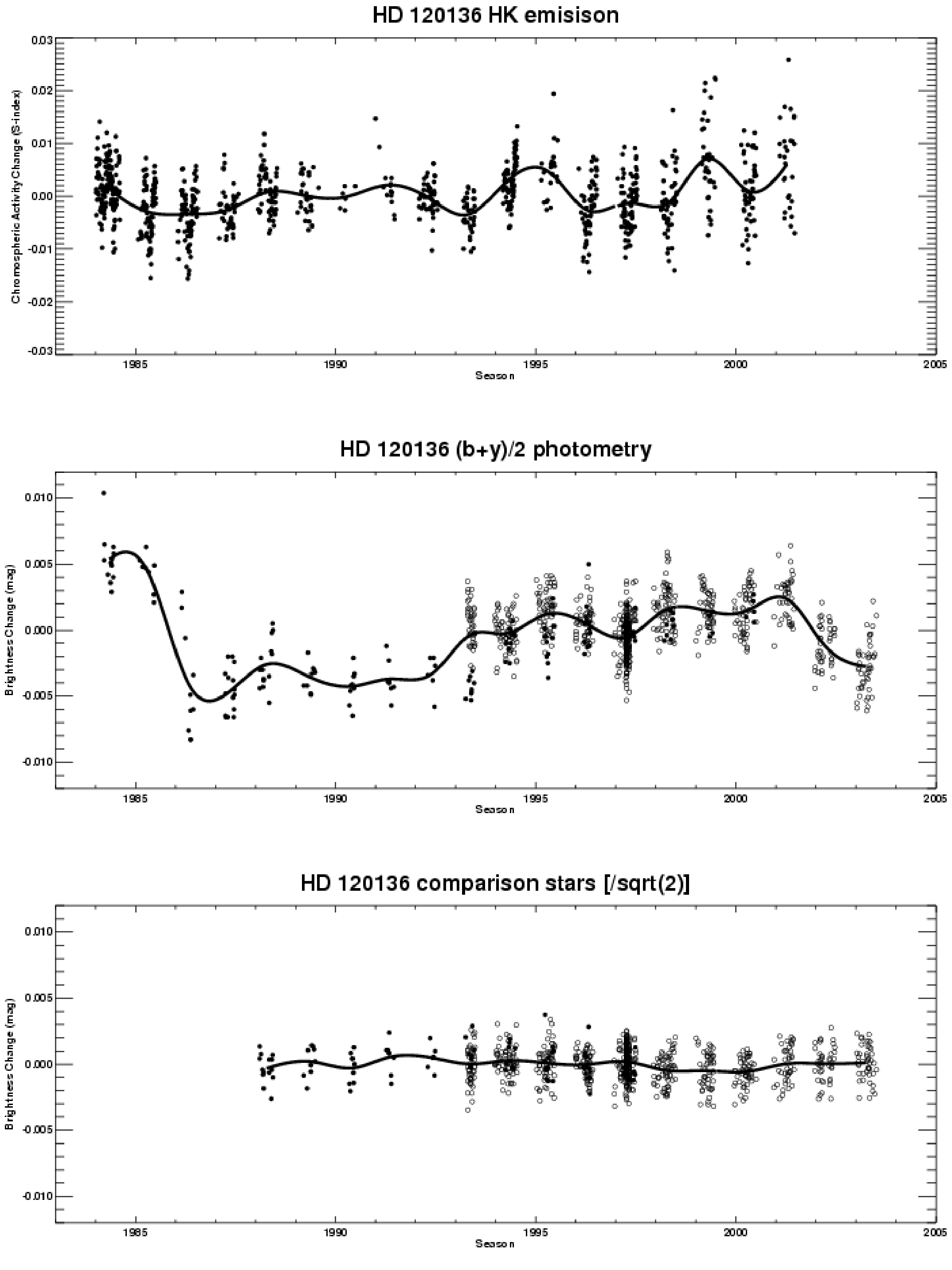

FIG. 3l—HD 120136

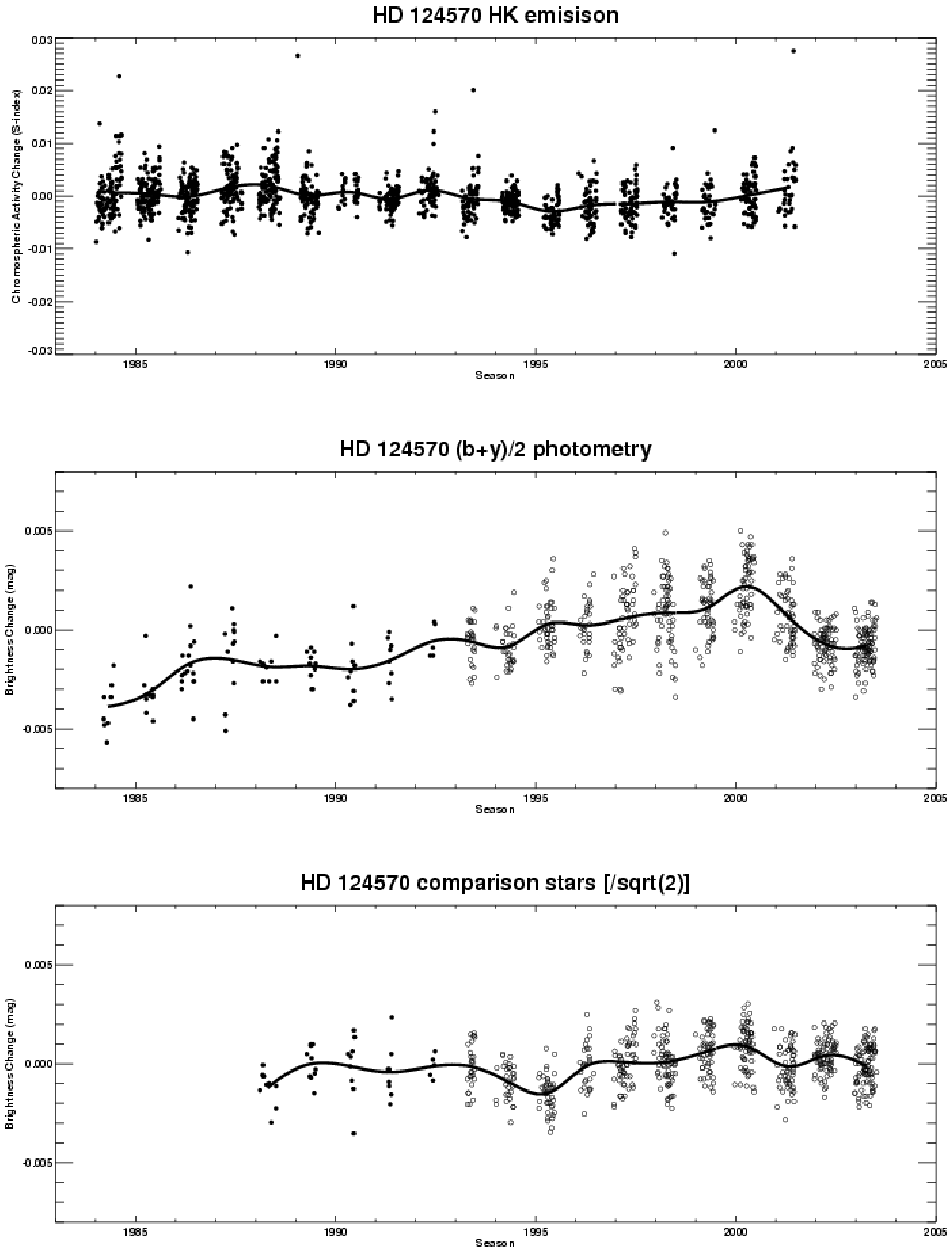

FIG. 3m—HD 124570

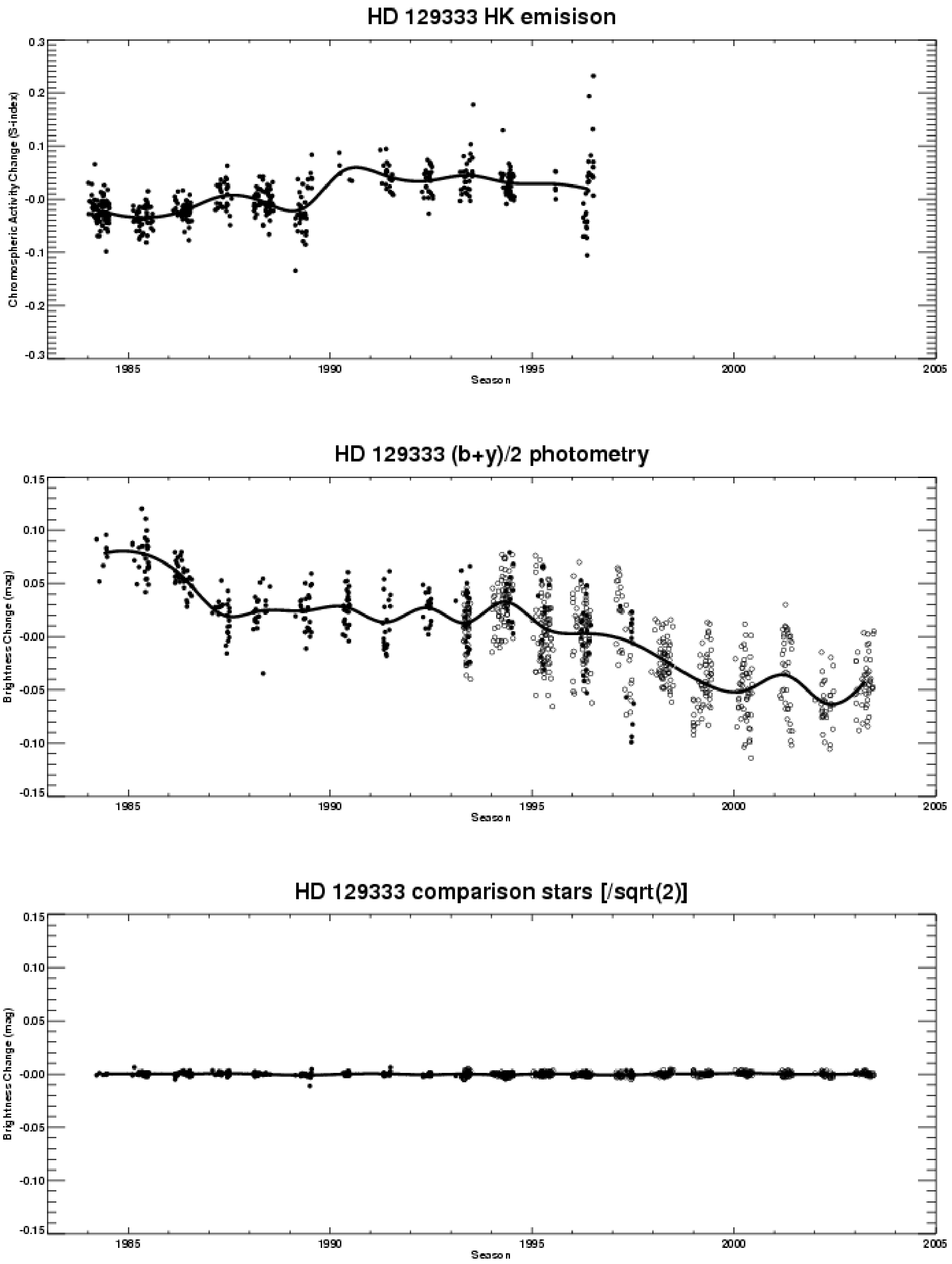

FIG. 3n—HD 129333

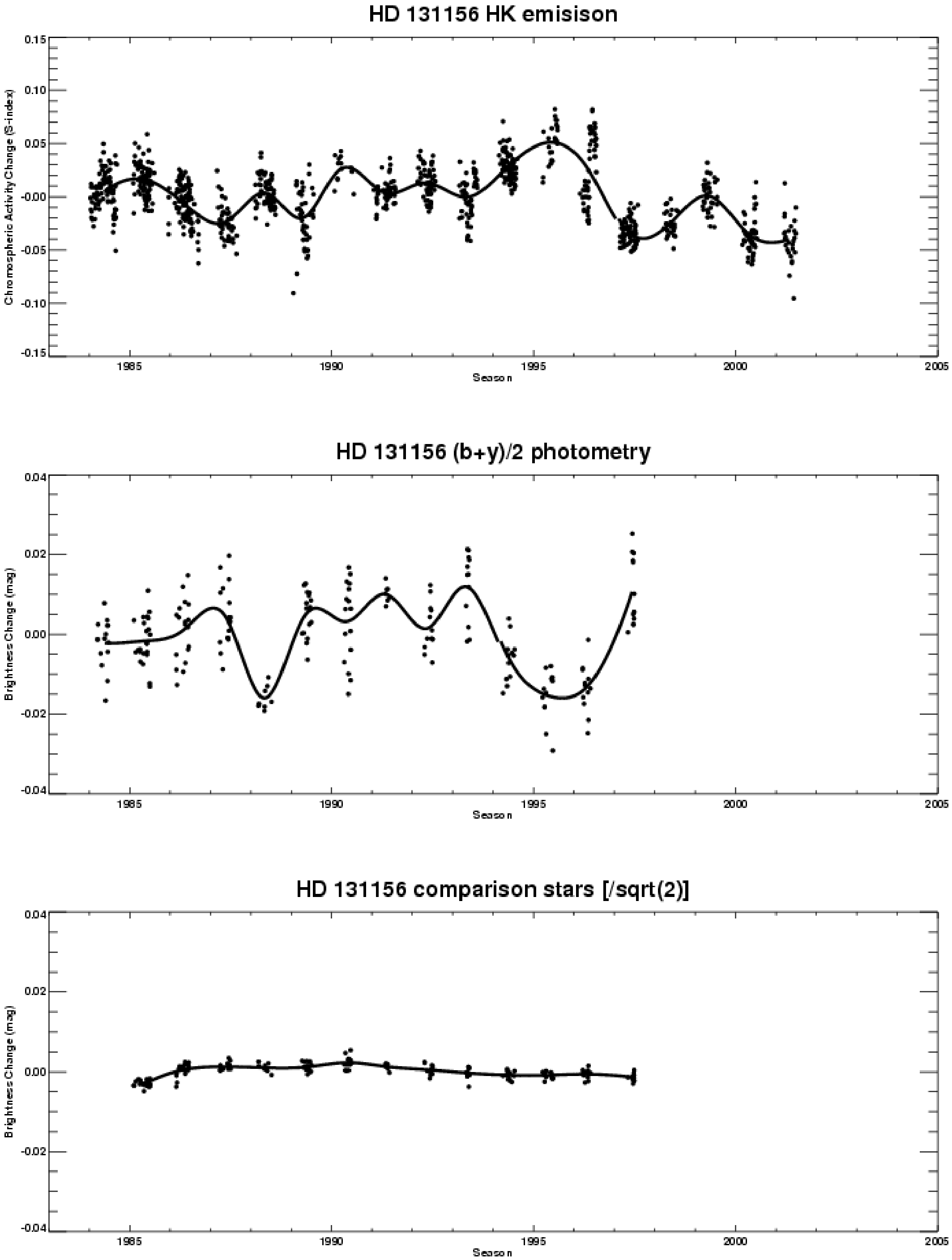

FIG. 3o—HD 131156

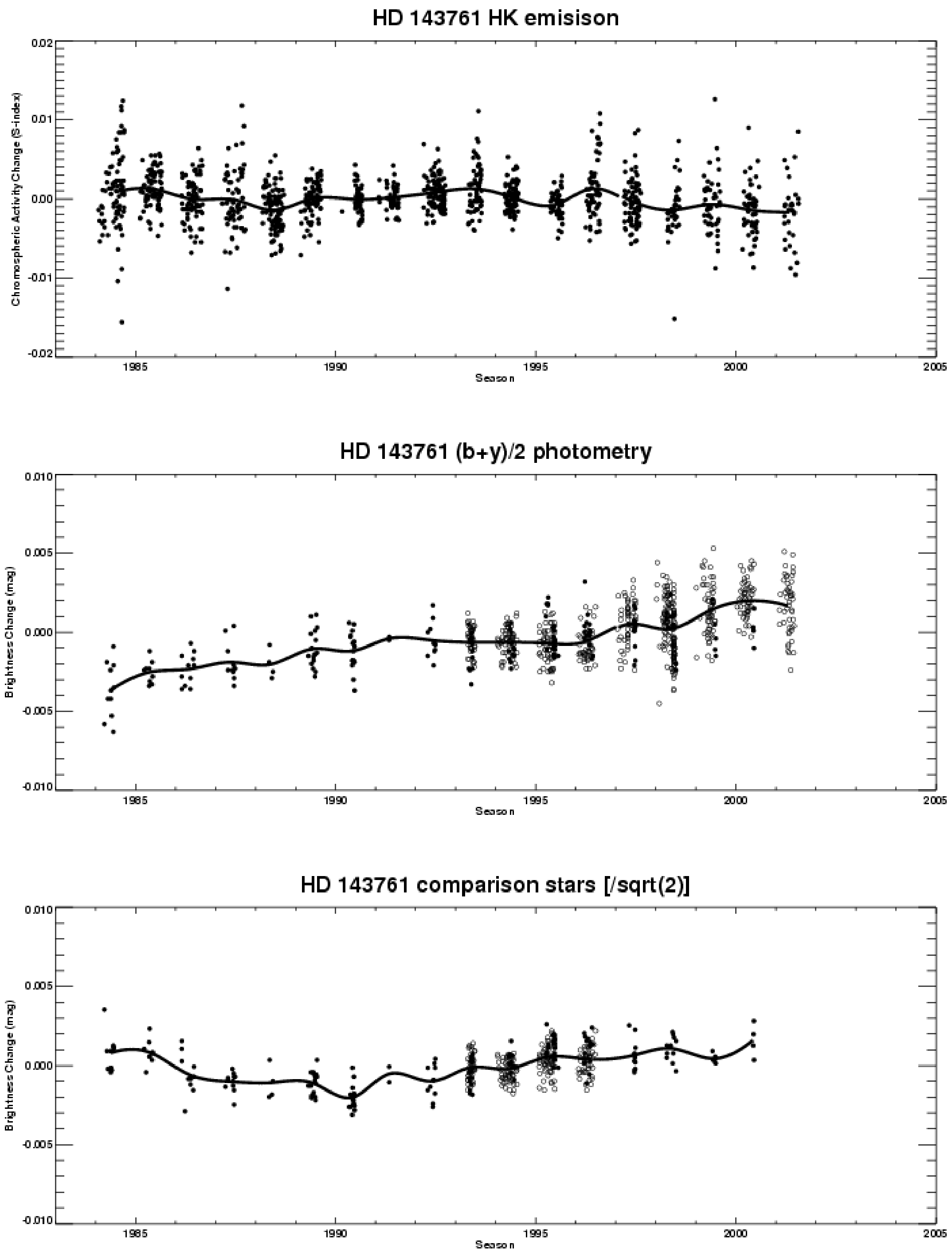

FIG. 3p—HD 143761

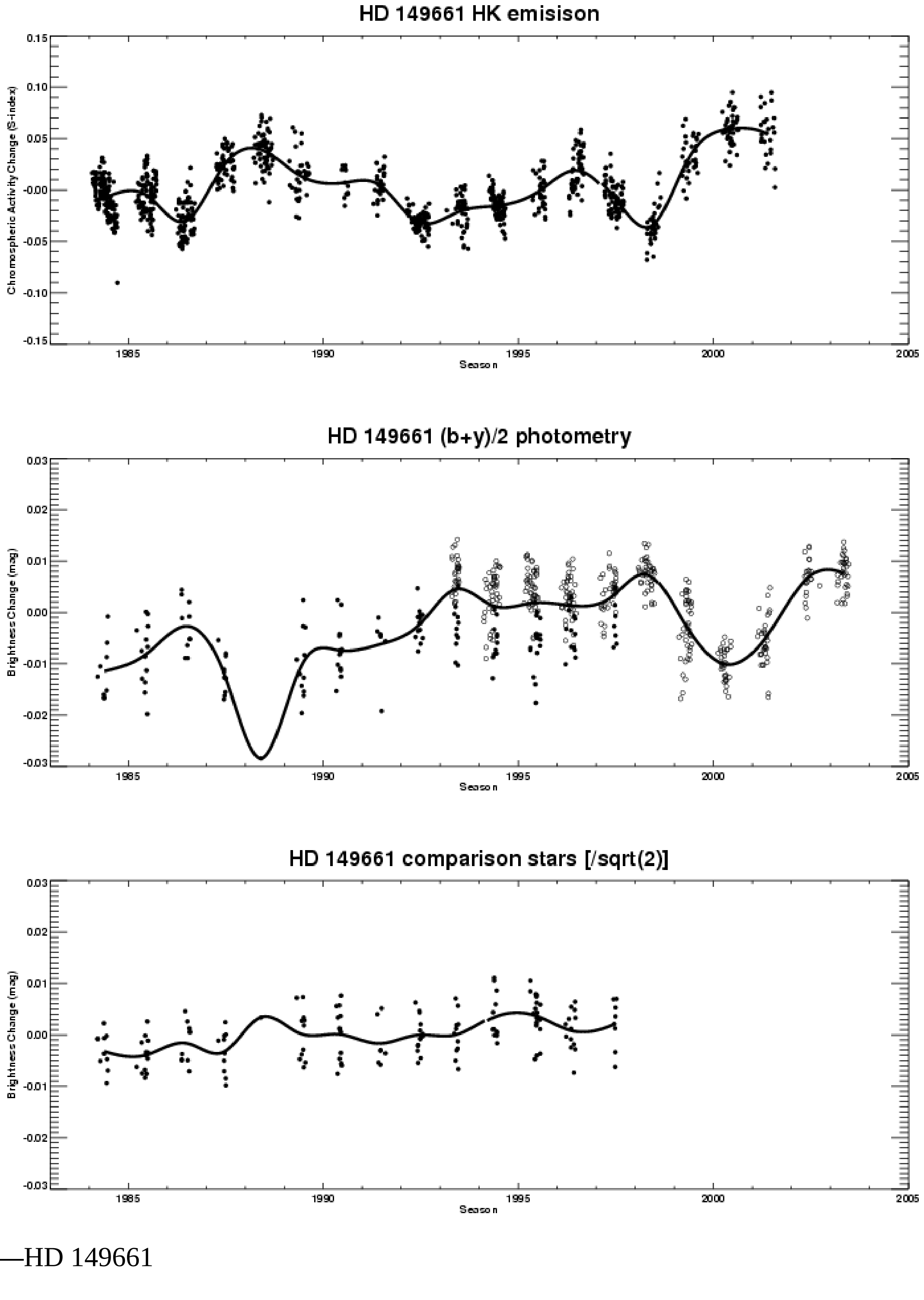

FIG. 3q—HD 149661

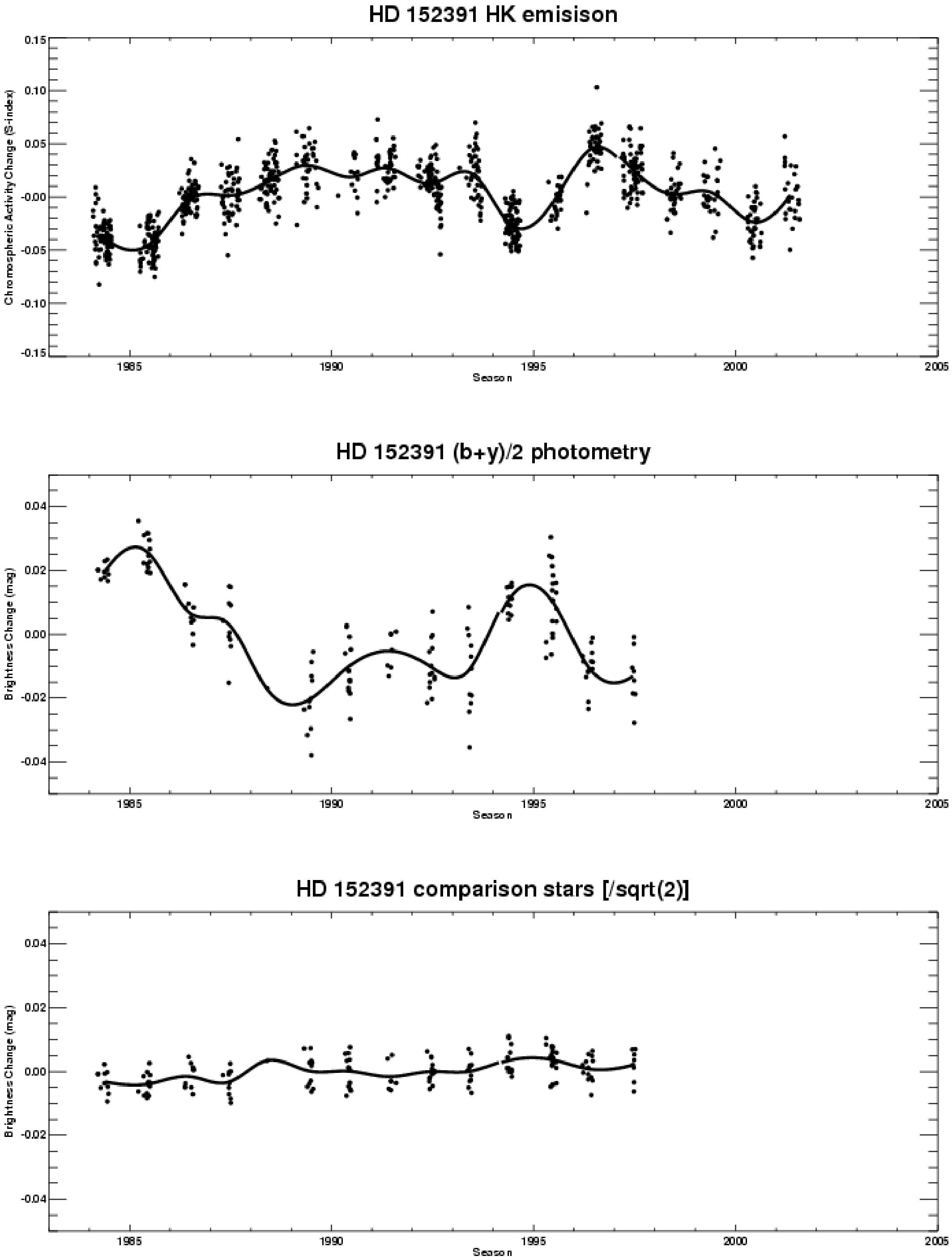

FIG. 3r—HD 152391

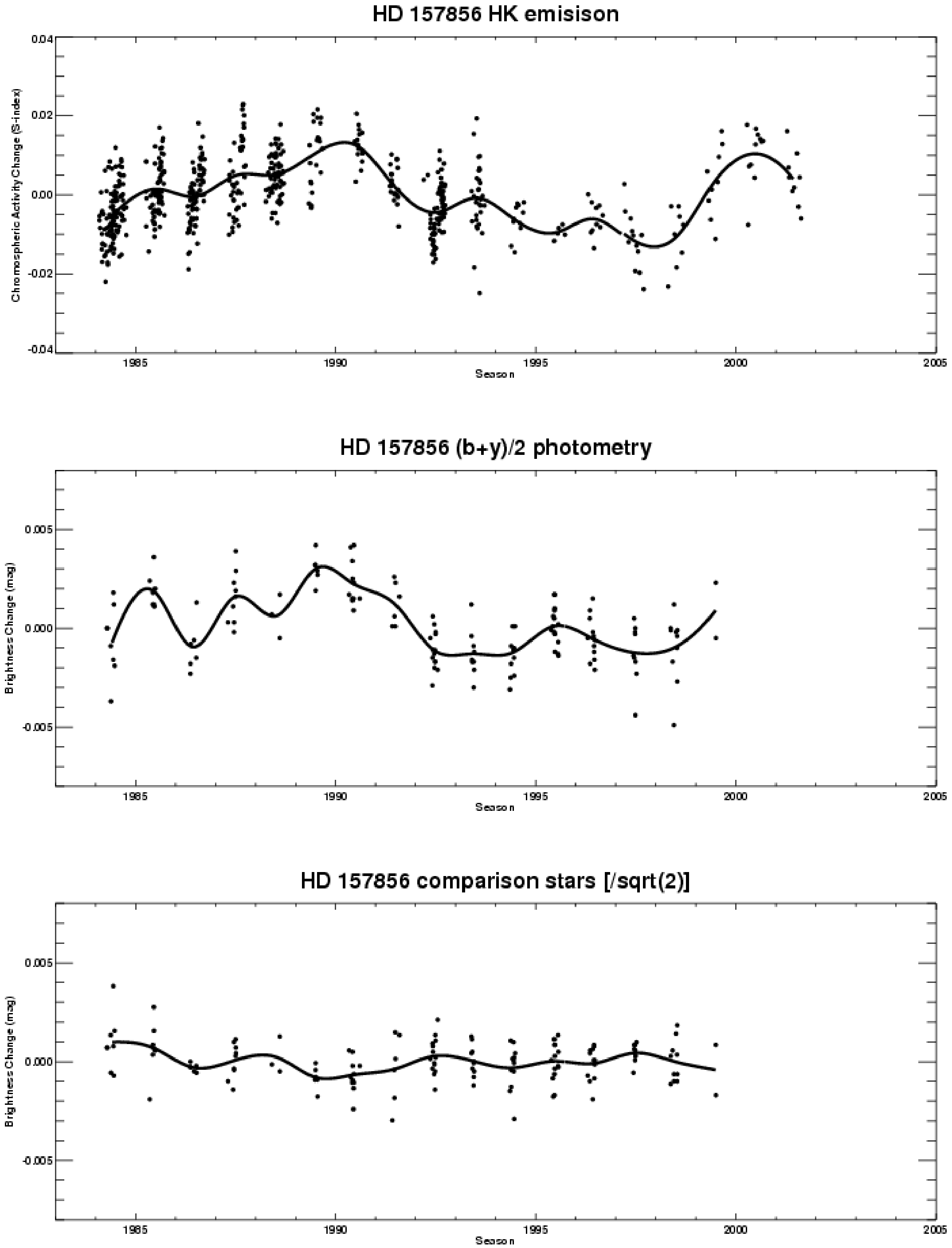

FIG. 3s—HD 157856

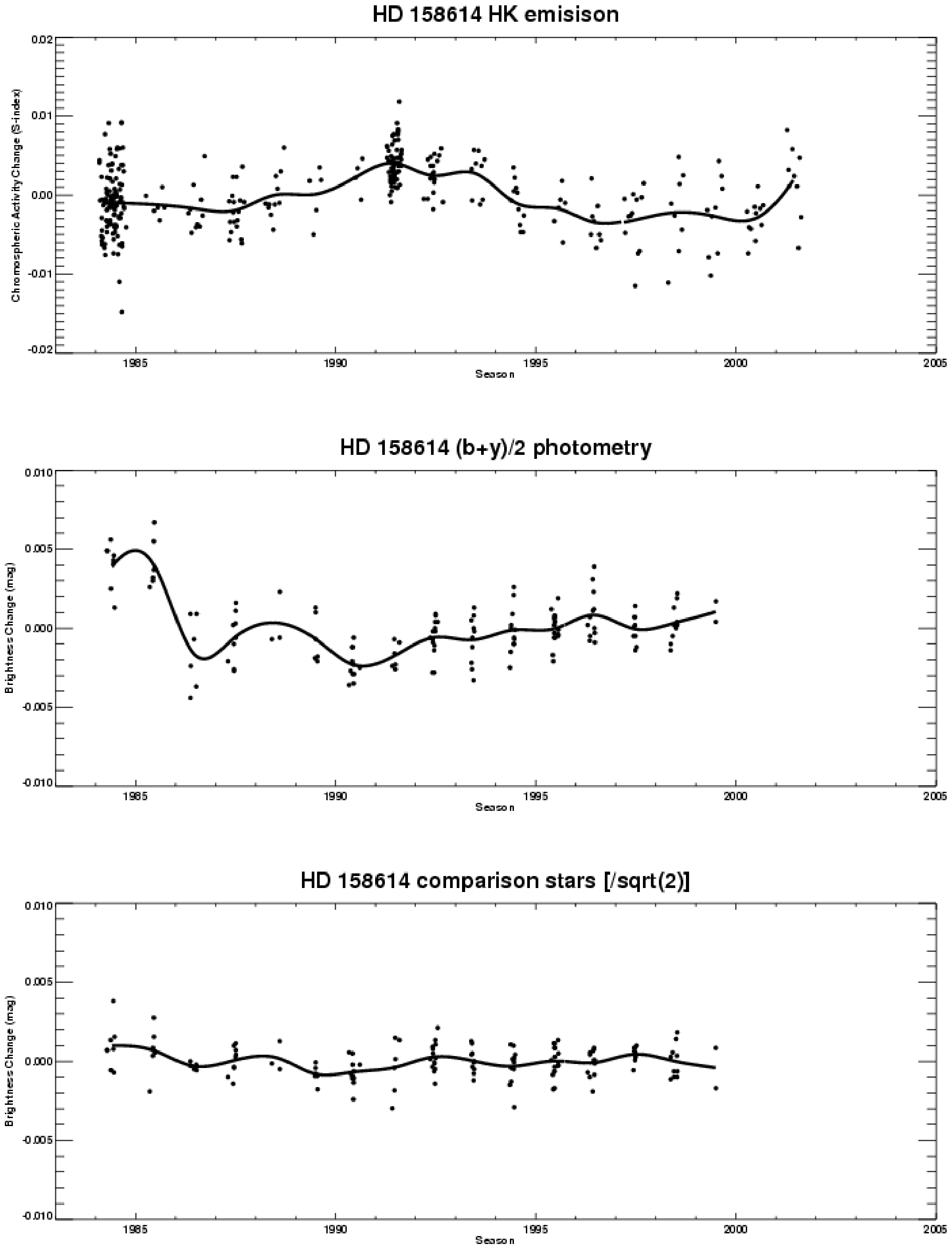

FIG. 3t—HD 158614

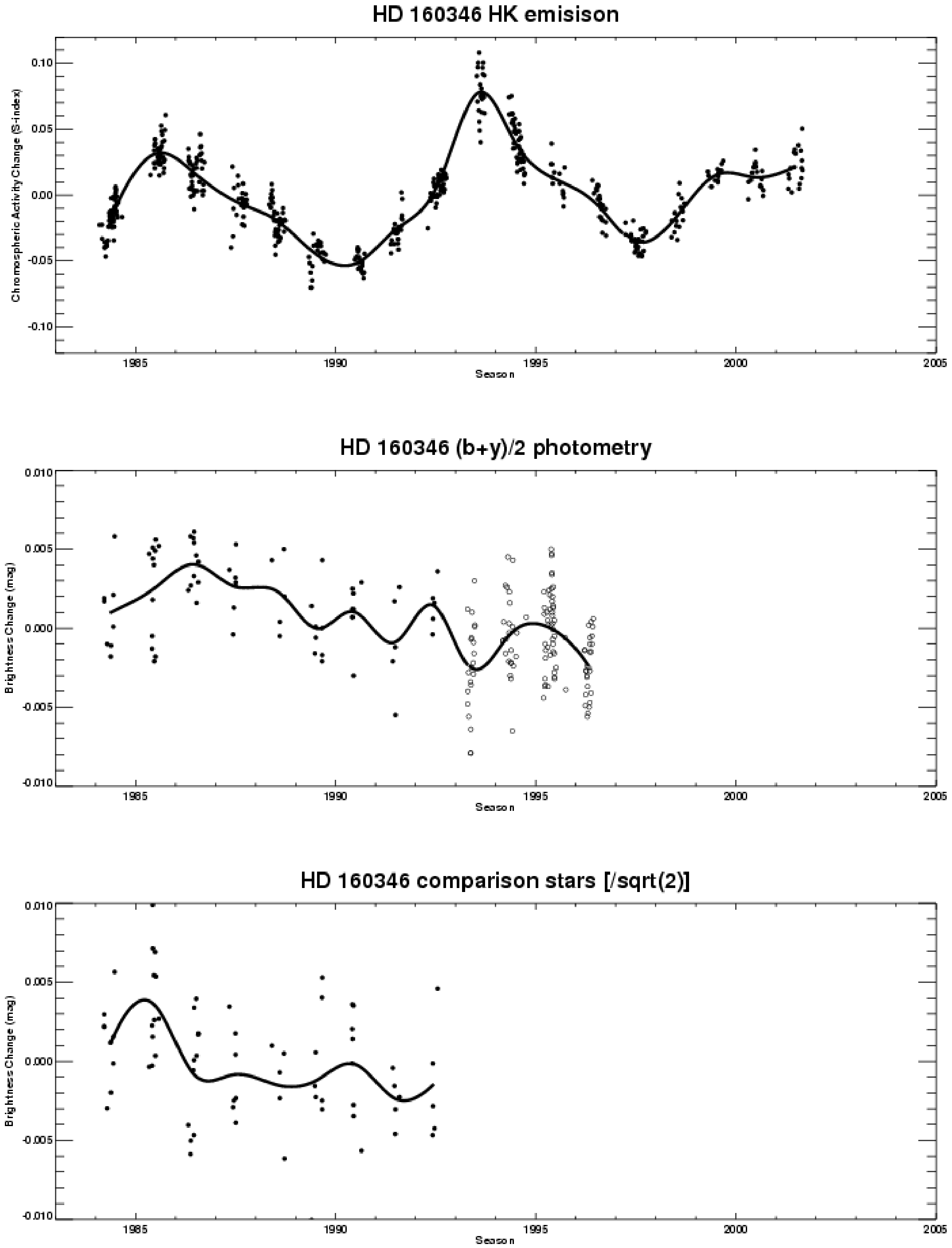

FIG. 3u—HD 160346

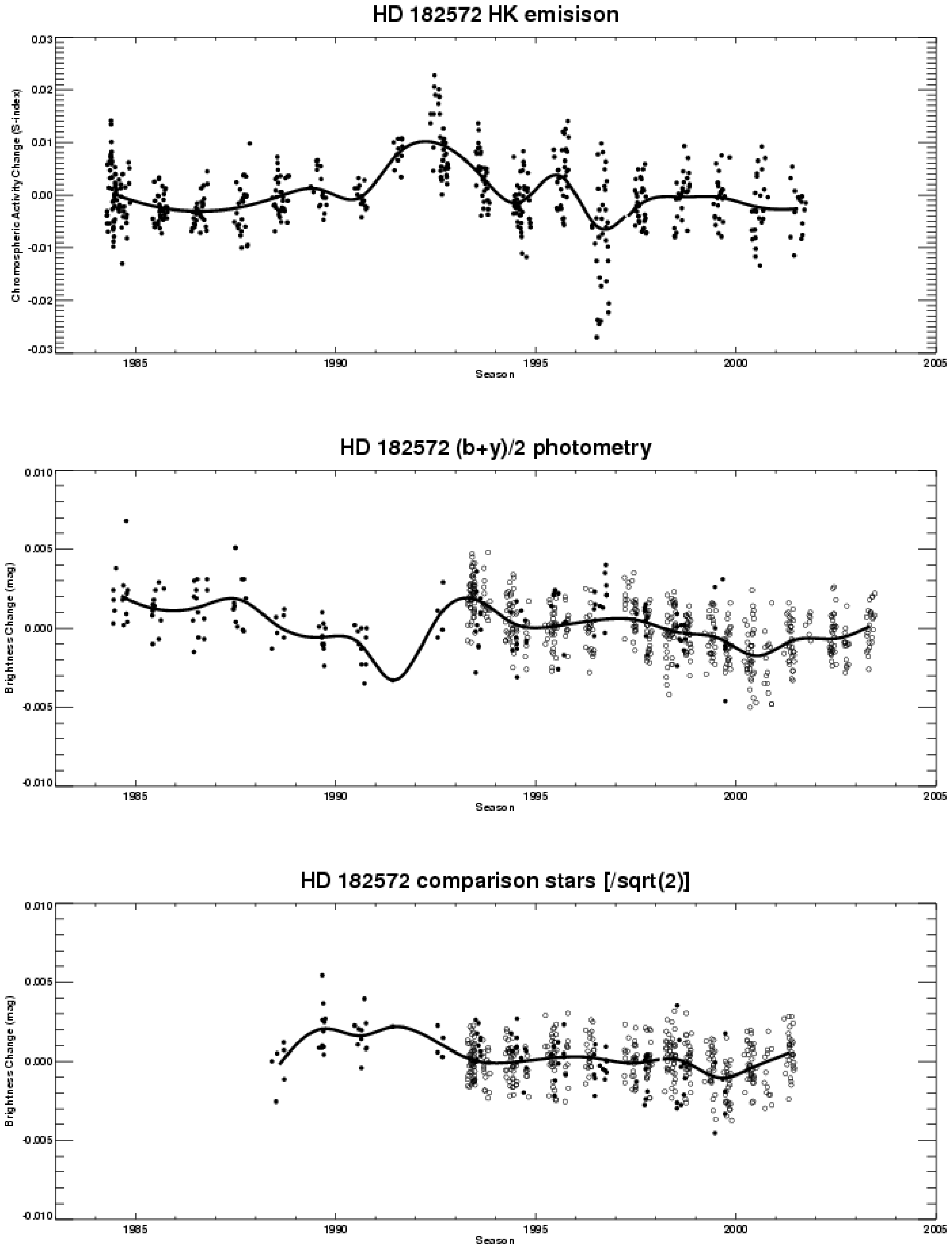

FIG. 3v—HD 182572

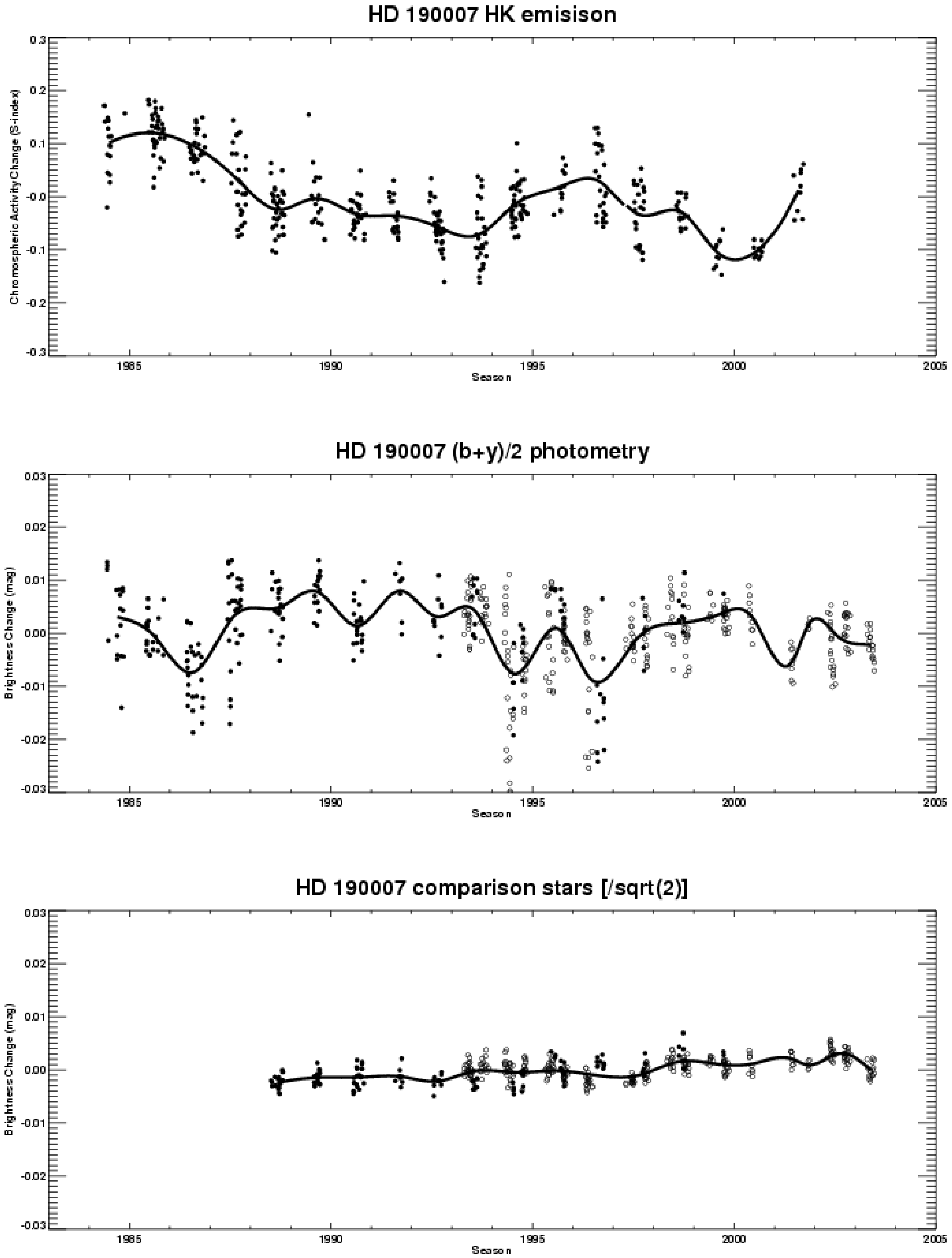

FIG. 3w—HD 190007

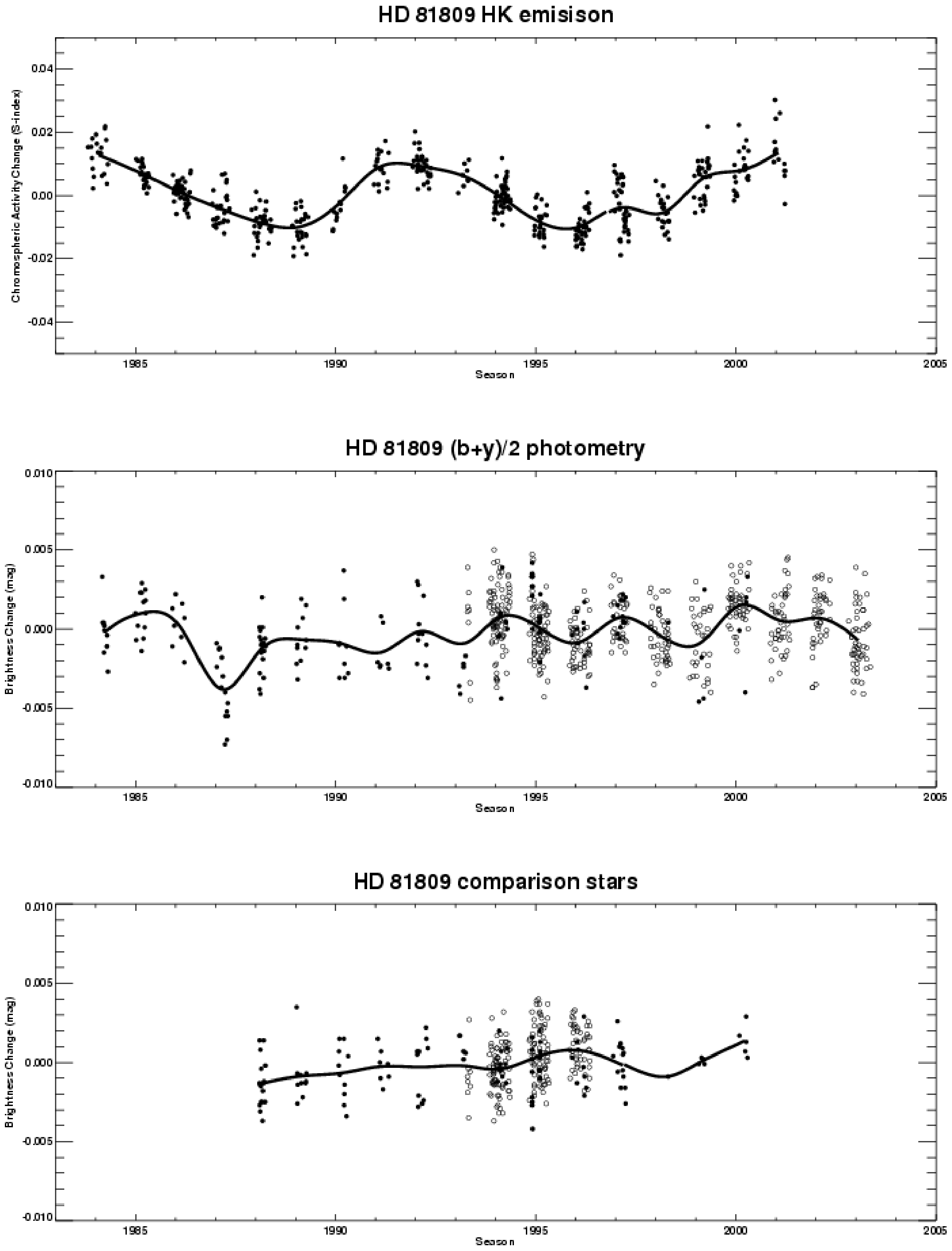

FIG. 4a—HD 81809. Chromospheric Ca II HK emission (*upper*), photometric program star (*middle*) and photometric comparison star (*lower*) time series plots for the stars of our sample having only one valid comparison stars. Brightness increases upward in all cases, and the bottom panel is included merely to show that the illustrated comparison star pair is unsuitable.

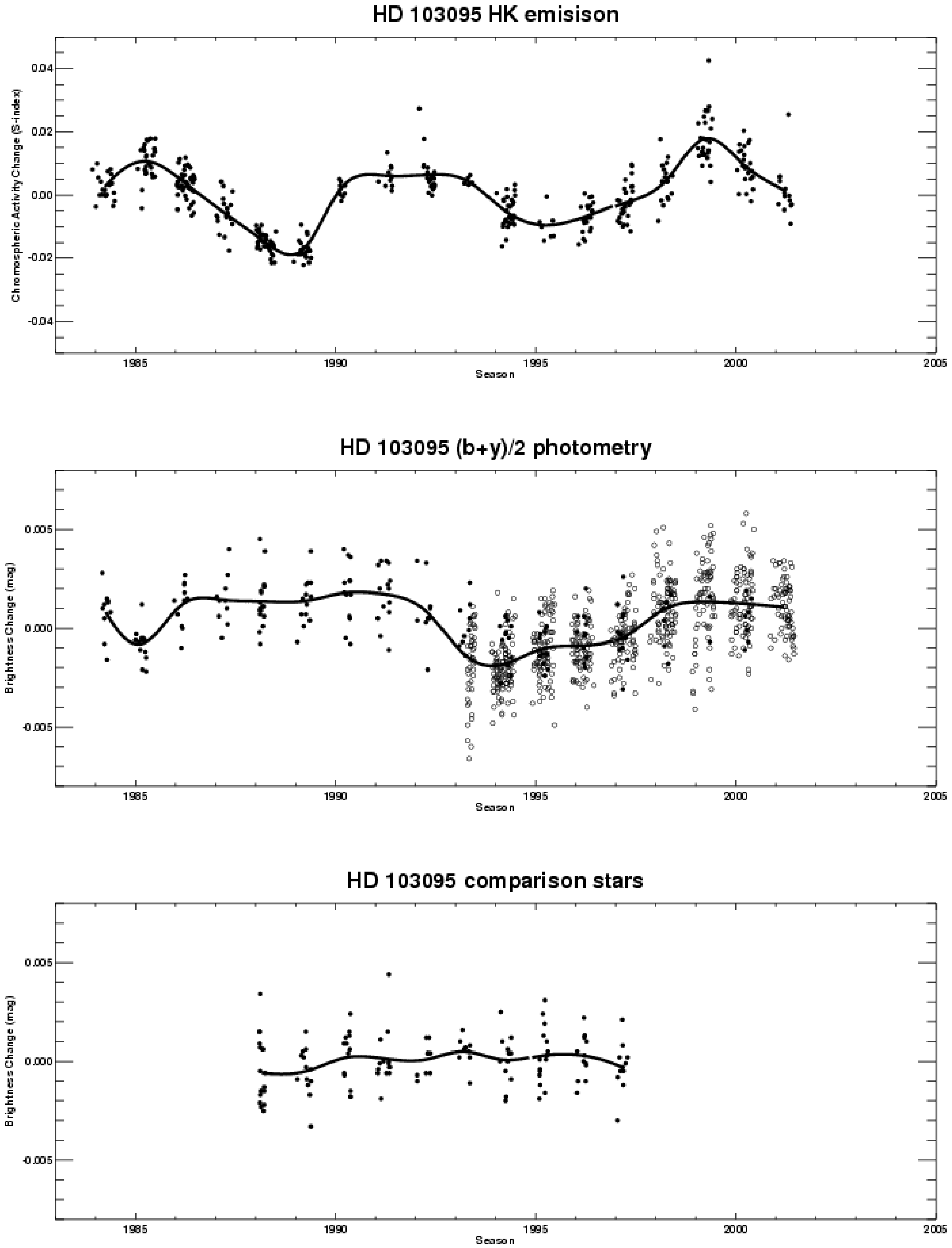

FIG. 4b—HD 103095

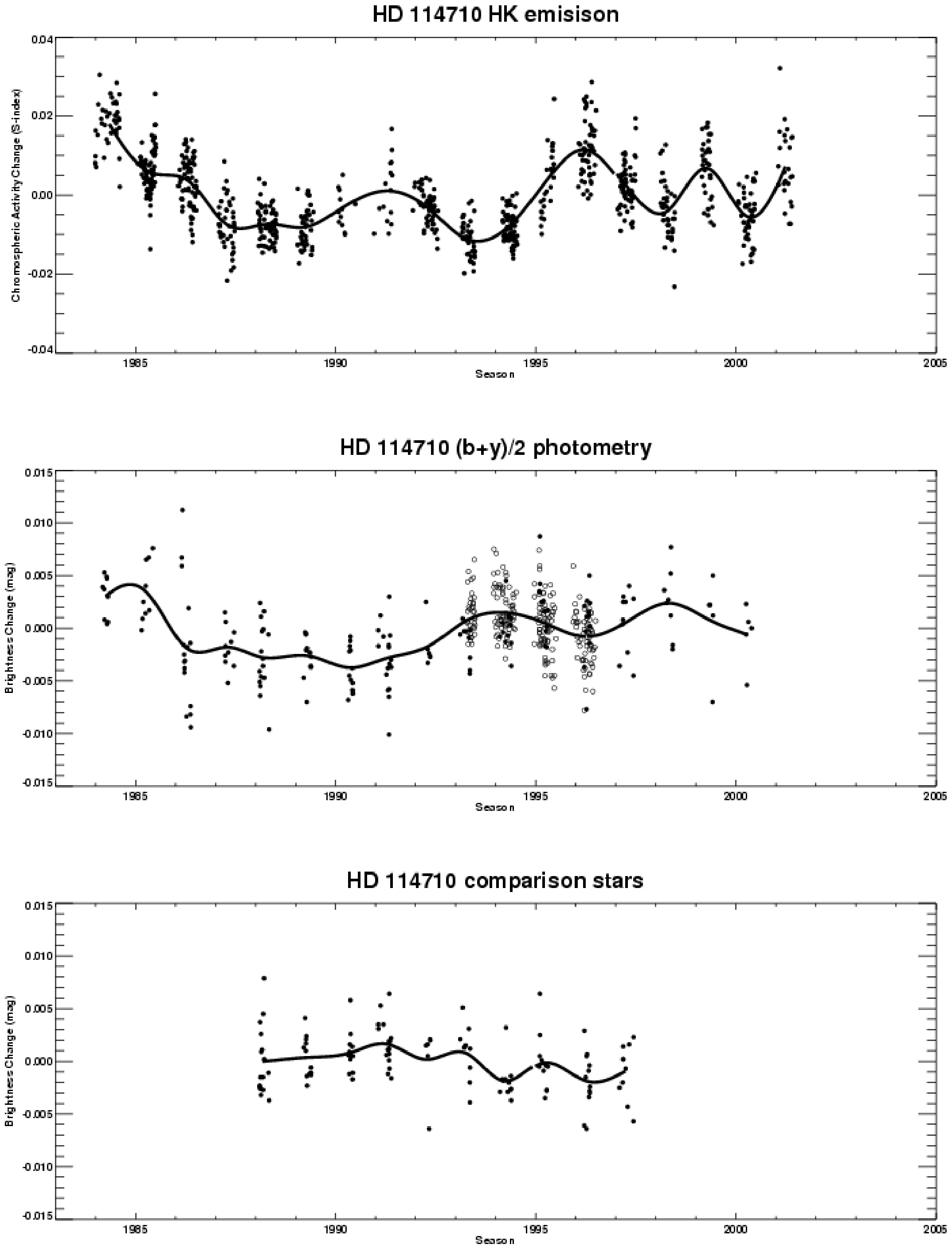

FIG. 4C—HD 114710

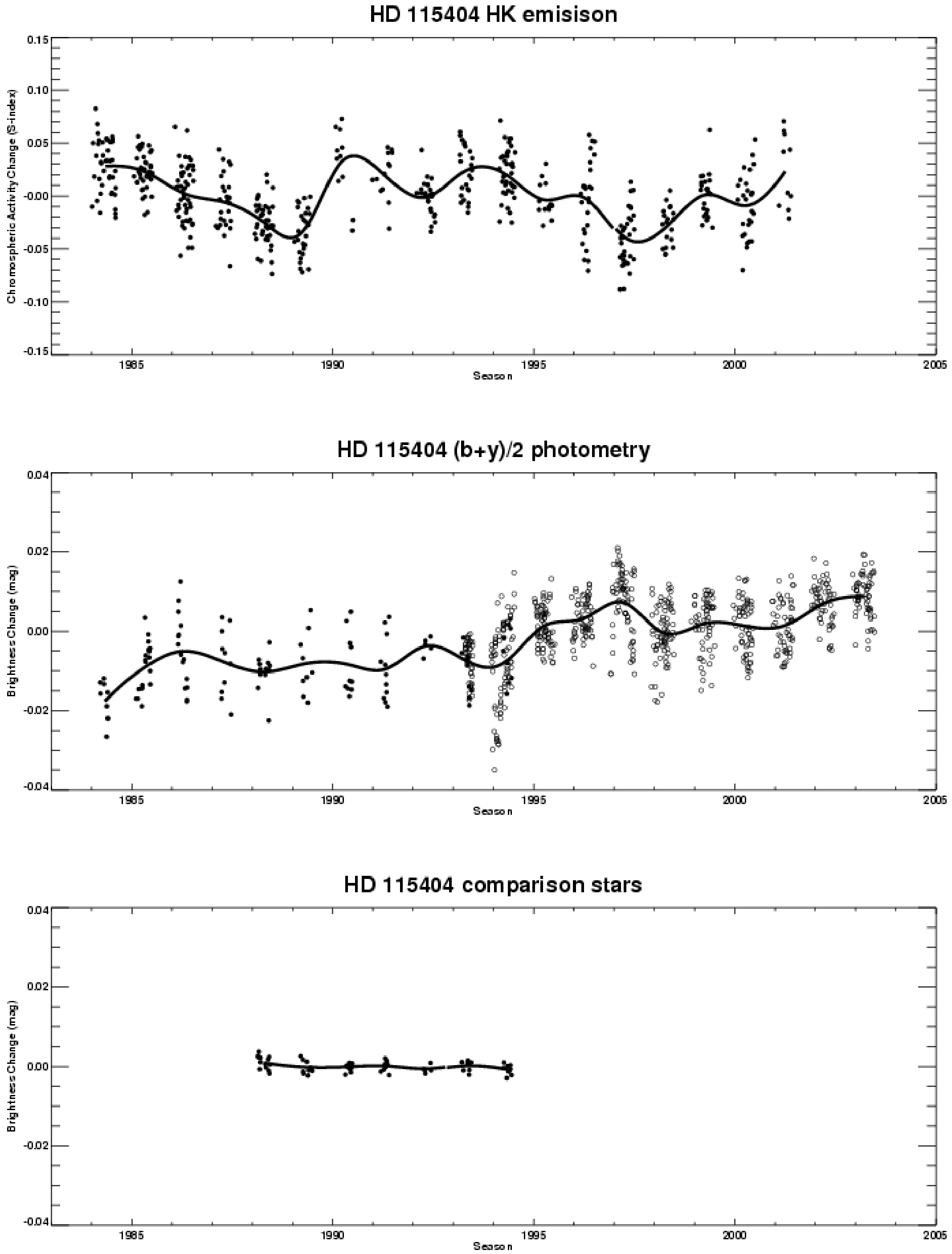

FIG. 4d—HD 115404

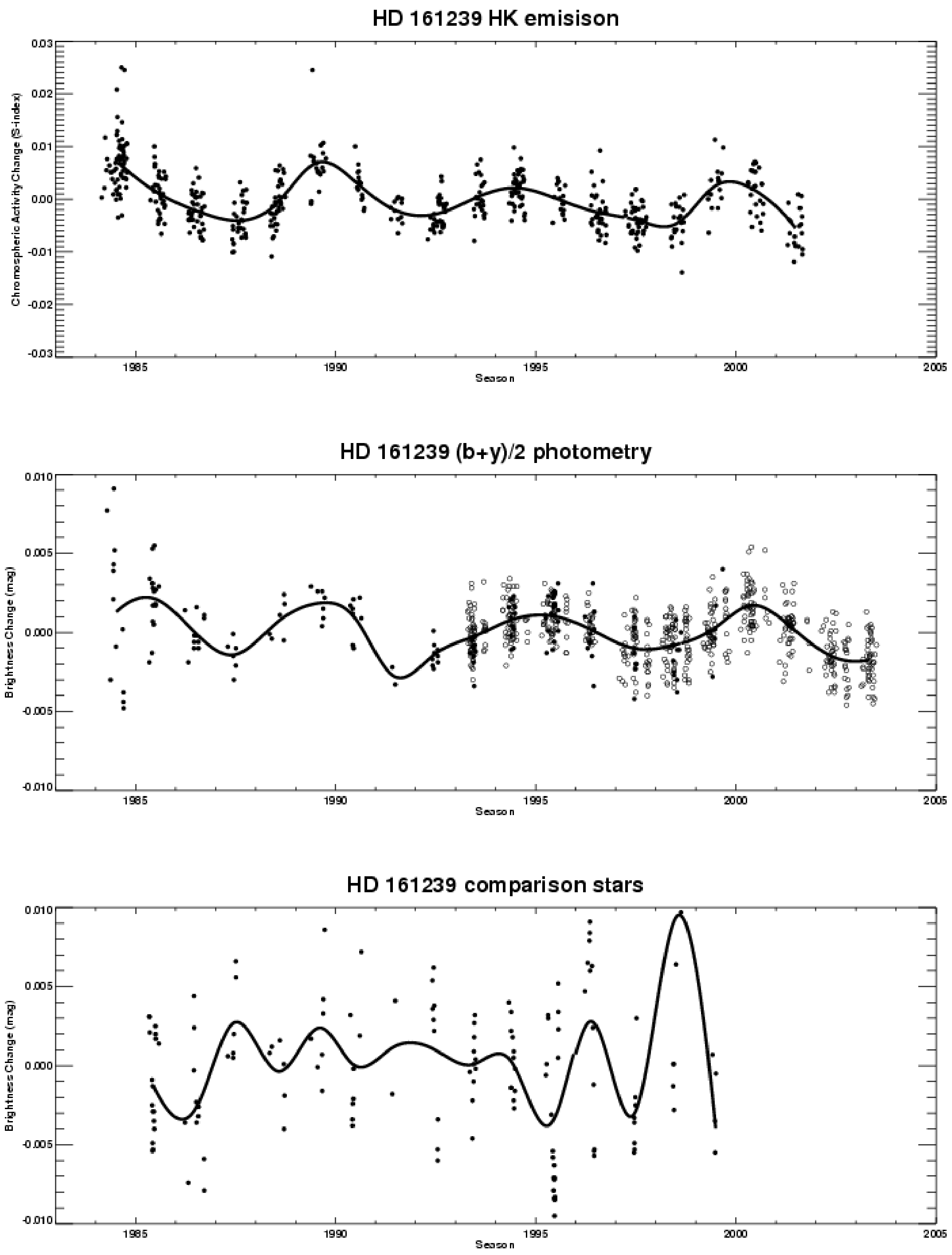

FIG. 4e—HD 161239

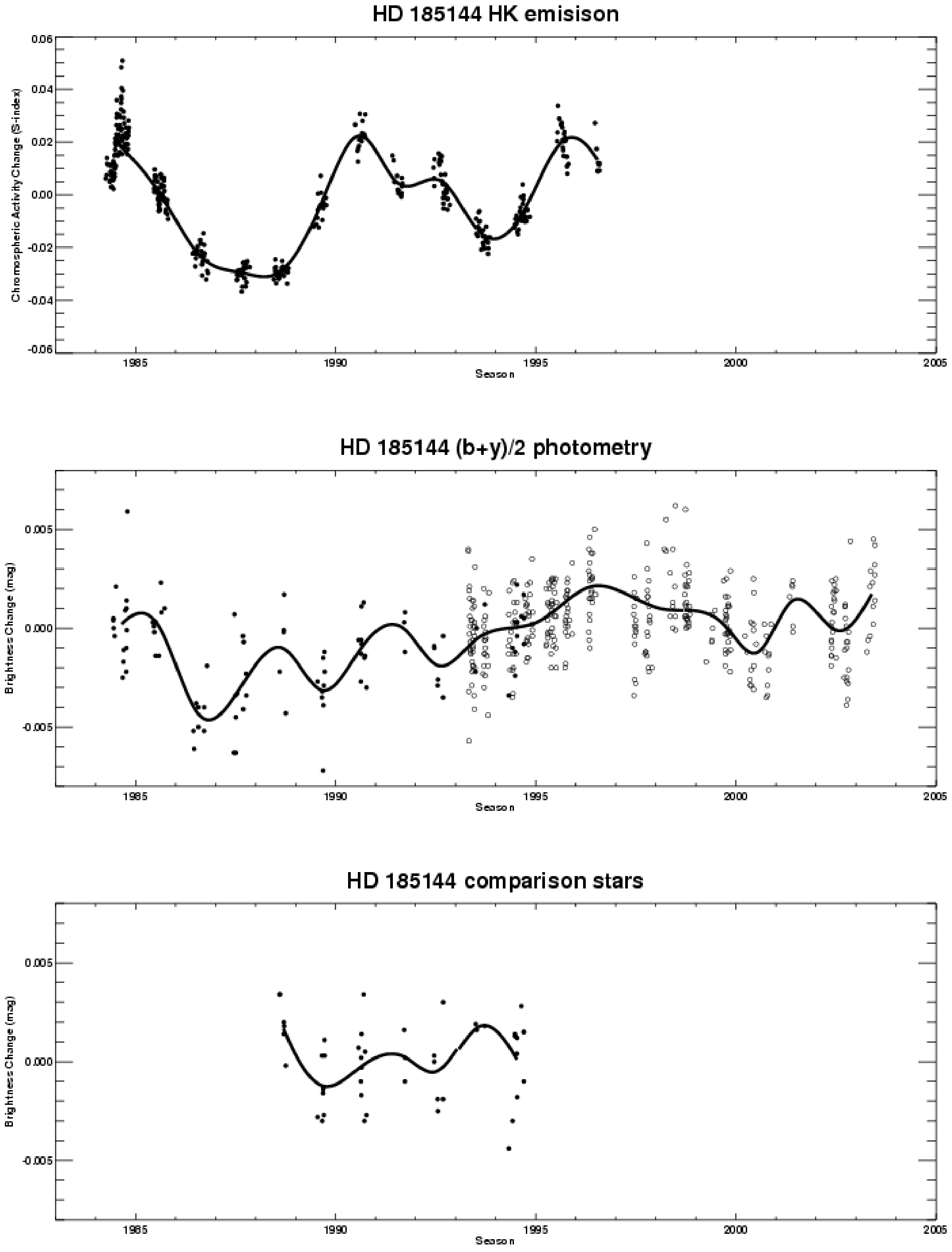

FIG. 4f—HD 185144

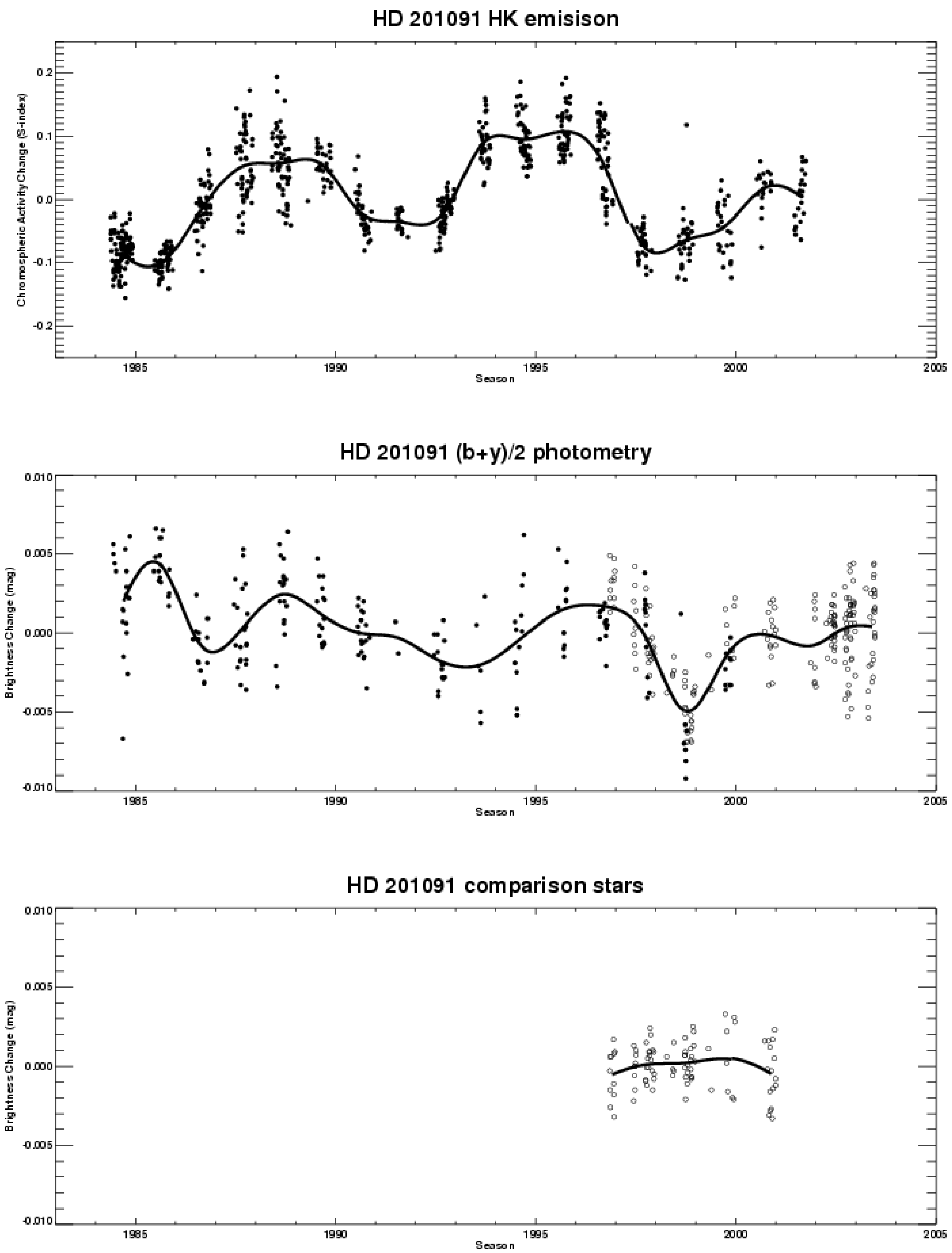

FIG. 4g—HD 201091

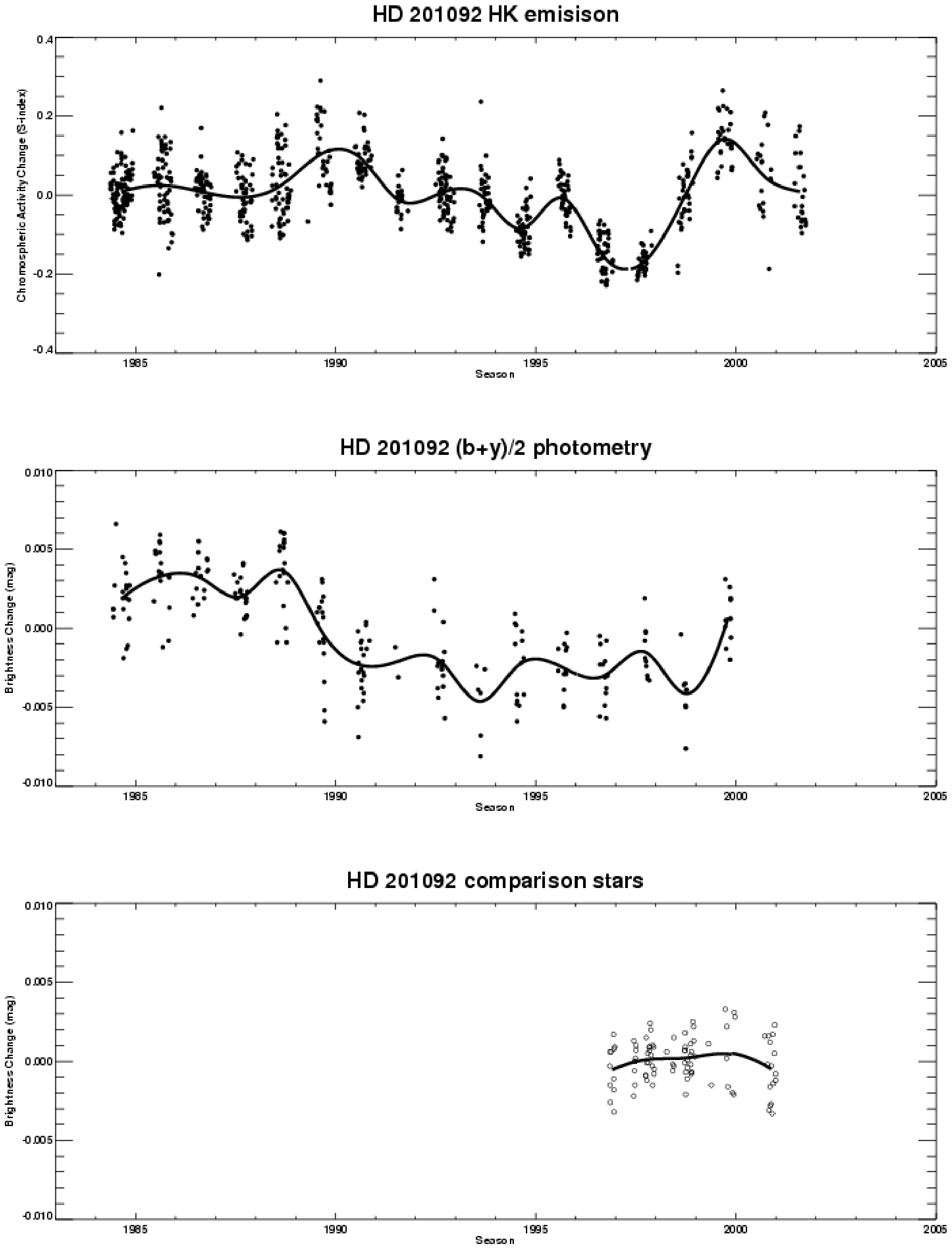

FIG. 4h—HD 201092

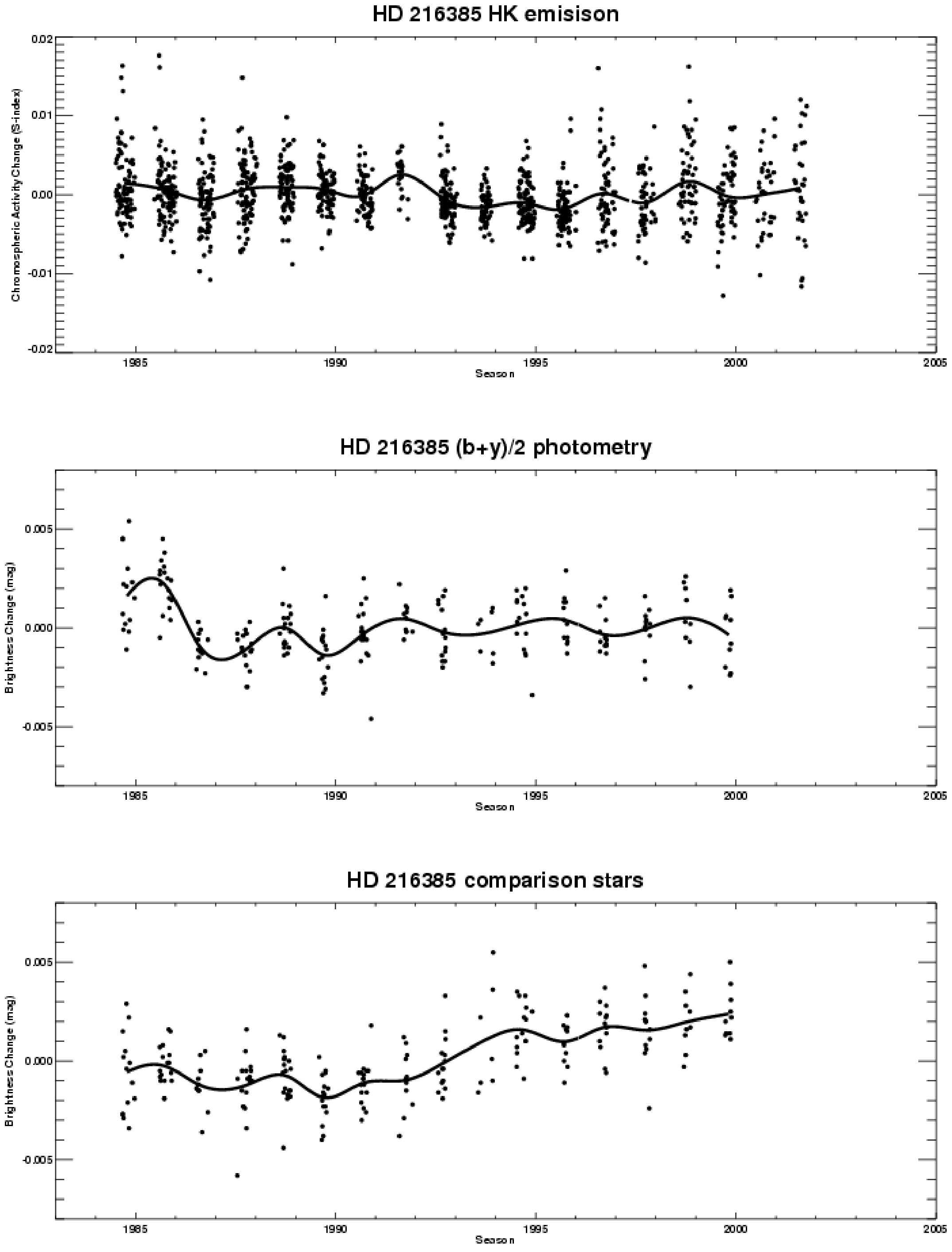

FIG. 4i—HD 216385

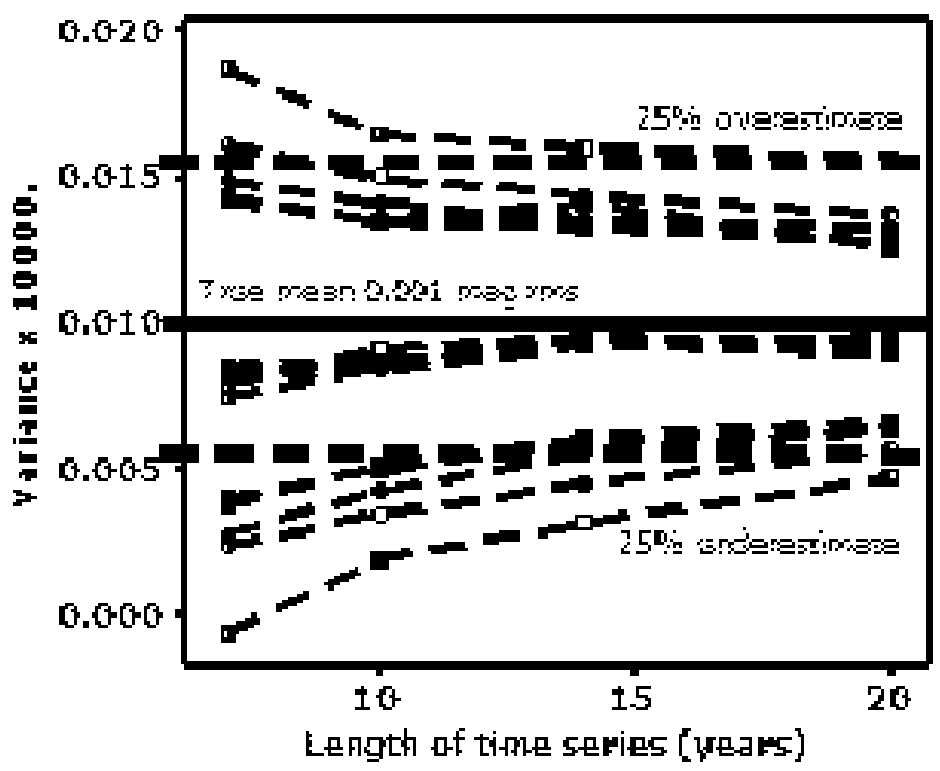

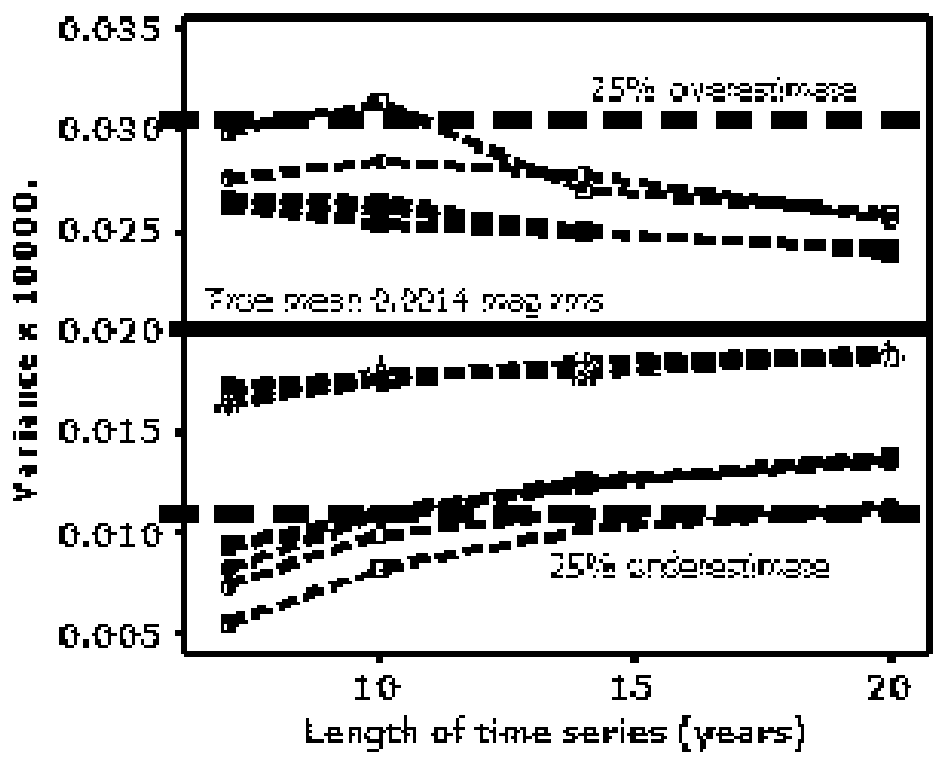

FIG. 5.—*upper*: Simulation for a program star with 0.001 mag rms intrinsic variation. Upper quartile, median, and lower quartile values are connected by lines over time for a comparison star pair having 0.00035, 0.0005, and 0.0007 mag rms variation (*filled squares*), 0.0010 mag variation (*open circles*), and 0.00140 mag variation (*open squares*). The vertical scale is variance x 10000.The dashed lines indicate values of the variance corresponding to 25% over- and under-estimates of the standard deviation. The smaller the comparison star intrinsic variability, the more likely it is that the calculated program star variability will lie within a 25% error band. *lower*: Simulation for a program star with 0.0014 mag rms intrinsic variation.

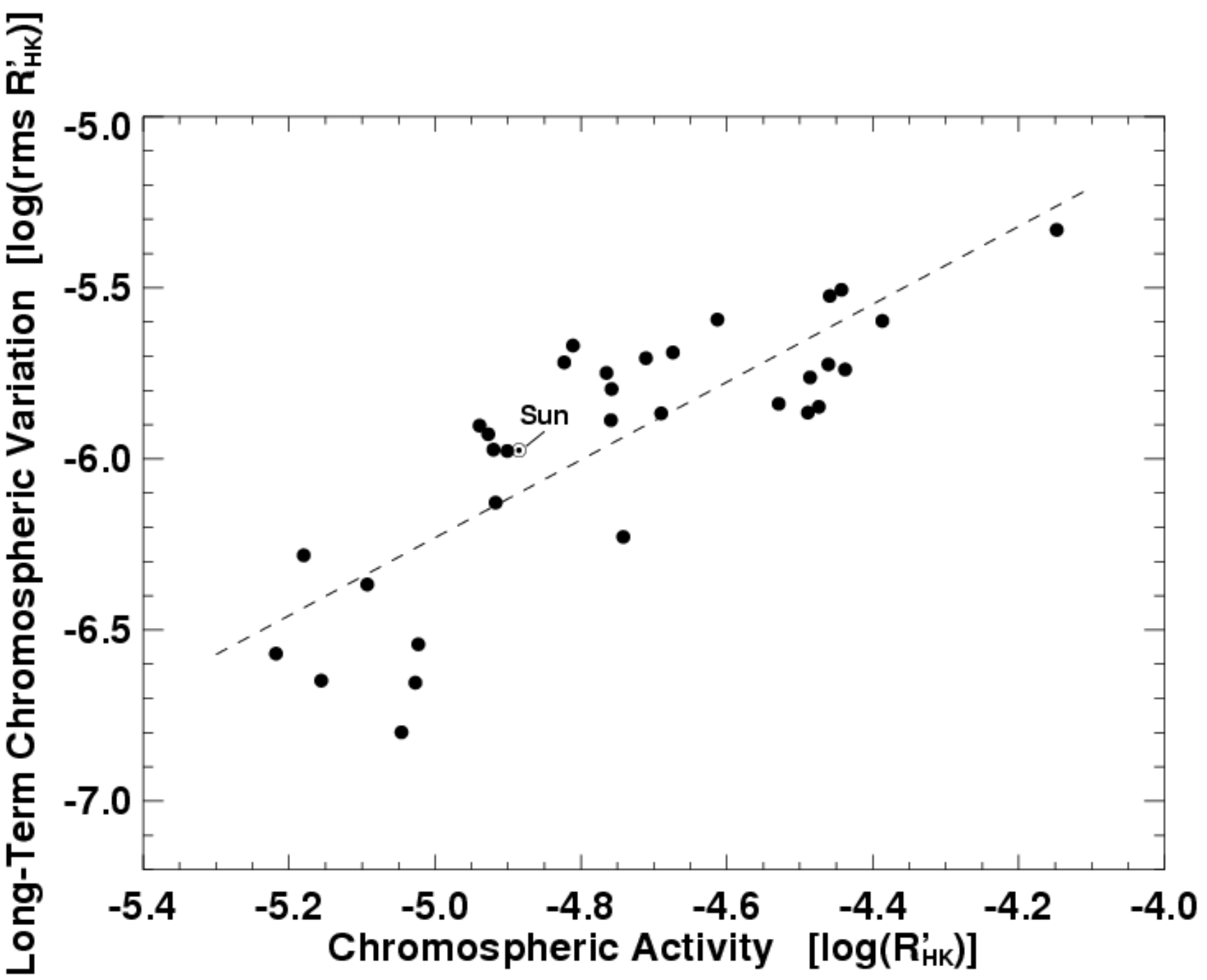

FIG. 6—Long-term (cycle time scale) chromospheric variation vs. average chromospheric activity level.

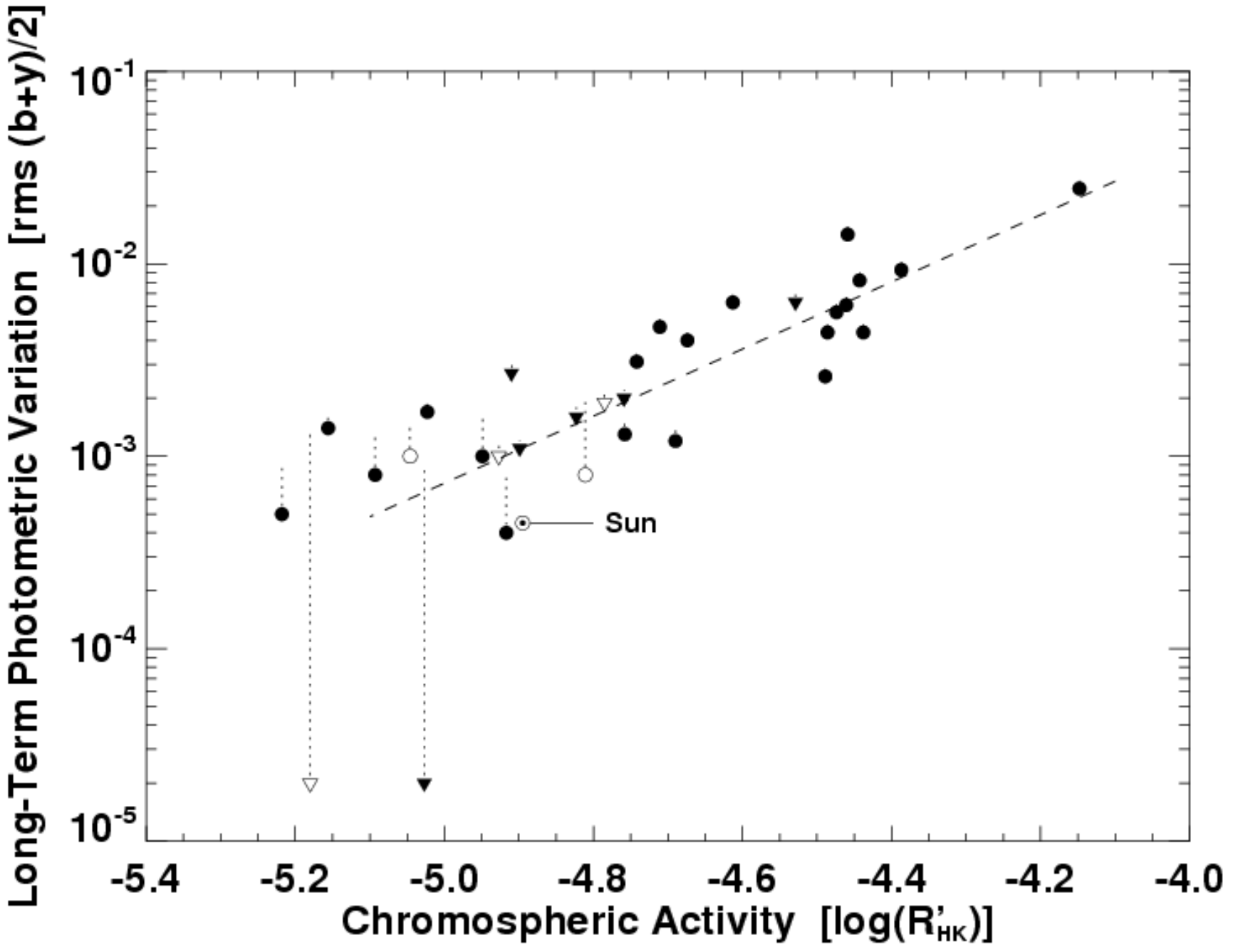

FIG. 7— Long-term (cycle time scale) photometric variation vs. average chromospheric activity level.

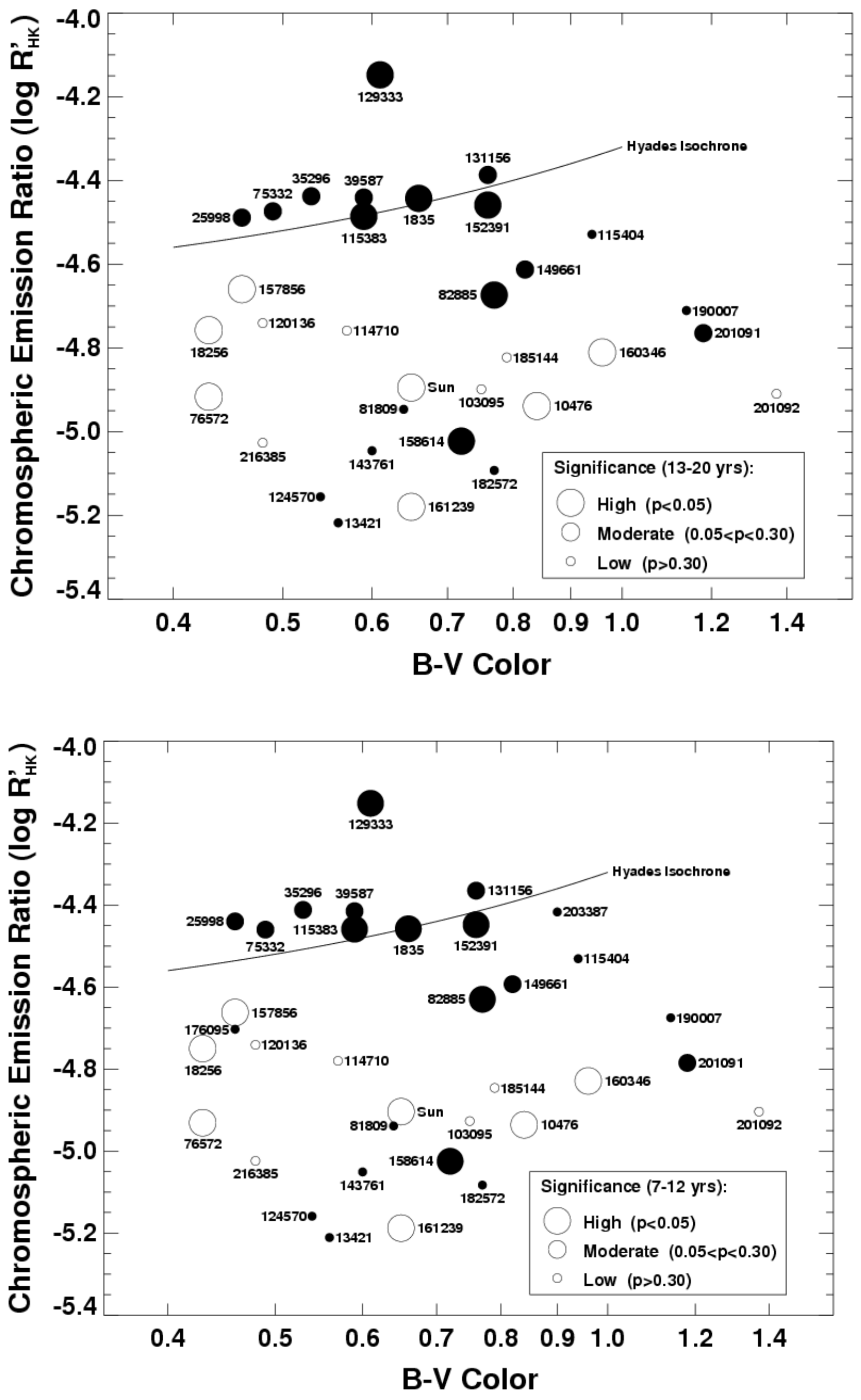

FIG. 8— Correlation between photometric brightness and HK emission variations for long timescales based on 13–20 years of observation. (*upper*) and 7-12 years of observation from Paper II. (*lower*). Many correlations are strengthened and none of the 32 surviving stars in the longer sample show reversal in the sense of the correlation.

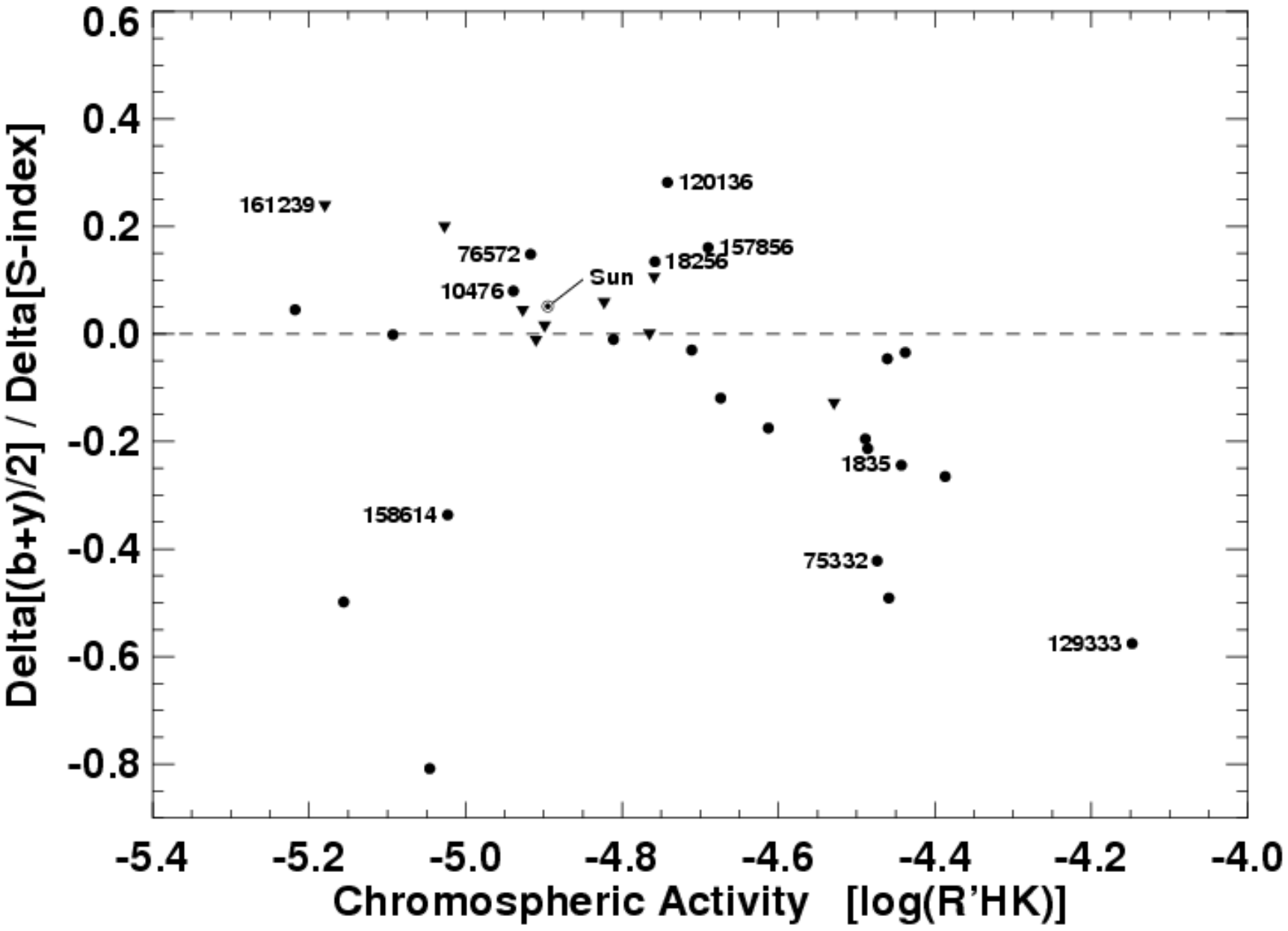

FIG.9—Slope of the regression of photometric brightness variation on HK emission variation, plotted as a function of average chromospheric level.